\documentclass[12pt,preprint]{aastex}

\def\kms{\relax \ifmmode {\,\rm km\,s}^{-1}\else \,km\,s$^{-1}$\fi}
 
\def\degree{\mbox{$^{\circ}$}}
\def\Mso{{M$_{\rm \odot}$}}
\def\mlr{\rm M$_\odot$~yr$^{-1}$}
\def\cm3{${\rm cm}^{-3}$}

\shorttitle{AGB stars interacting with the ISM}
\shortauthors{Villaver, Garc\'{\i}a-Segura \& Manchado} 

\begin{document}

\title{The Interaction of Asymptotic Giant Branch Stars with the Interstellar Medium}
 
\author{Eva Villaver\altaffilmark{1}, Arturo Manchado\altaffilmark{2,3} \& Guillermo
  Garc\'{\i}a-Segura\altaffilmark{4}}

\altaffiltext{1}{Departamento de F\'{\i}sica Te\'orica, Universidad
  Aut\'onoma de Madrid, Cantoblanco 
28049  Madrid, Spain; {\tt eva.villaver@uam.es}}  

\altaffiltext{2}{Instituto de Astrof\'{\i}sica de Canarias, V\'{\i}a
        L\'actea S/N, E-38200 La Laguna, Tenerife, Spain} 

\altaffiltext{3}{Consejo Superior de Investigaciones Cient\'{\i}ficas,
  Spain. {\tt amt@ll.iac.es}} 

\altaffiltext{4}{Instituto de Astronom\'{\i}a-UNAM, Apartado postal 877,
       Ensenada, 22800 Baja California, M\'exico
 {\tt ggs@astrosen.unam.mx}}
\begin{abstract}
We study the hydrodynamical behavior of the gas expelled by moving
Asymptotic Giant Branch Stars interacting with the ISM. Our models follow the wind
modulations prescribed by stellar evolution calculations, and we cover
a range of expected relative velocities (10 to 100 \kms), ISM
densities (between 0.01 and 1 \cm3), and stellar
progenitor masses (1 and 3.5\Mso). We show how and when bow-shocks, and cometary-like
structures form, and in which regime the shells are subject to
instabilities. Finally, we analyze the results of the simulations in
terms of the different kinematical stellar populations expected in the Galaxy.

\end{abstract}


\section{INTRODUCTION}

At the end of their lives low-and intermediate-mass stars (those with main
sequence masses between 1-8 \Mso) 
ascend the Asymptotic Giant Branch (AGB) in the HR diagram,
where, one the most remarkable characteristic of their
evolution is the ejection of the stellar envelope in a series of high
mass-loss rate events during the thermal-pulsing phase at the end of the
AGB stage. 

While evolving, stars move within the gravitational field of the
galaxy, and AGB stars are no exception. The association of a star with the
different galactic components determines on a first order its velocity
within the Galaxy. This stellar motion influences the structure and 
dynamics of the ejected AGB envelope as the stellar mass-loss interacts with
the local  Interstellar Medium (ISM)  \citep{Vgm03}. Moreover, mostly due to
ram pressure stripping, an important fraction of the mass ejected by the star
along the AGB is left downstream of the motion, forming cometary-like
structures behind the star, even when low velocity interactions are
considered (see e.g. \citealt{Vgm03}).    

The striking observations of the Mira AB binary system \citep{Metal07}
confirmed this theoretical scenario revealing a
surrounding arclike structure and a stream of material stretching
2\degree~away in opposition to the arc. Recently, more AGB stars showing the
effects of the interaction with the ISM in their circumstellar envelopes have
been found (e.g. \citealt{Lib08,Lib09,Lib10,Mr07,Uetal10,Jor11}). 

AGB stars eventually become Planetary Nebulae (PNe), and as the stellar effective
temperature increases, the AGB envelope becomes ionized.  In PNe the
asymmetries developed as a 
consequence of the interaction process with the ISM are in many cases a major
morphological feature (see
e.g. \citealt{Tk96,Xetal96,Bth93,Tmn95,Sz97,Gvm98,Chu09,Rl09,Ran08,Sze03,Bet11}).  

\cite{Gur69} was the first to suggest the interaction of the PN with
the ISM  as a possible mechanism to explain the observed asymmetries.  The first 
theoretical studies (see e.g. \citealt{Smi76,Isa79,Bss90,Sbs91})
arrived at the conclusion  that the nebula fades away before any disruption of
the nebular shell becomes noticeable unless high relative velocities or
densities were involved. But these studies only
considered the interaction process once the nebular shell was already
formed. In \cite{Vgm03}, by studying the
interaction process as the star evolves along the AGB phase,
we demonstrated that the star-ISM interaction can be predominant even at low
ISM densities and/or velocities, given that it appears as a direct consequence 
of evolution of the evolving AGB winds. Other models to study the
interaction process in PN shells have been published since then to explain the morphologies
observed in individual objects \citep{Sze03,Vs05,Wai06,Wai07c}. In particular, the detailed 
observations of the Mira cometary structure \citep{Metal07} have motivated 
a wealth of elaborated theoretical work of the interaction \citep{Wai07,Retal08,Rc08,Esq10},
with a stellar wind and ISM parameters chosen to reproduce the observations of this 
particular object. 

\cite{Rc08} presented analytical predictions for the velocity of the 
material in the wake of Mira,
as a function of the distance to the stellar source, and compare with the 21cm observations 
of the system. \cite{Wai07} studied numerically the interaction between the Mira wind and the 
ISM by using an isotropic and constant AGB wind with a mass-loss rate of 
3$\times$10$^{-7}$ \mlr (variations of the mass-loss by factors of 3 times this value were also 
considered) and a velocity of 5 \kms in a 
3D grid in order to match the overall observed structure. A 3D domain was
adopted as well by \cite{Retal08} with similar parameters for the constant AGB wind 
(7.7$\times$10$^{-7}$ \mlr, 10 \kms for the mass-loss and velocity respectively) as those used
by \cite{Wai07}. But in this case, with the goal of reproducing the 
double shock structure observed in the cometary head a dependency with latitude is set for 
the stellar wind. The 2D adaptive mesh refinement 
work of \cite{Esq10} aim to reproduce the broad-head narrow-tail structure observations 
of Mira that previous simulations \citep{Wai07,Retal08} failed to reproduce. For that, they 
consider a changing ISM environment (a dense ISM turns into a low density rarefied medium 
with the conditions of the local bubble) and used the same wind parameters as those adopted 
by \cite{Wai07}. What all this numerical simulations have in common is that the wind 
and ISM parameters have been chosen to reproduce the observations of a particular object.

A more general case of the interaction under a broader range of conditions has been 
explored in the 3D simulations by \cite{War07}. \cite{War07} examined a range of 
velocities for the interaction between 25 and 125 \kms. However, the AGB wind was 
assumed to be constant within each simulation. Four different values of the 
AGB mass-loss rates were explored: 
5$\times$10$^{-7}$ (for a relative velocity of 25 \kms 
through a ISM with 0.01 \cm3), 5$\times$10$^{-6}$ (for 50, 75, and 125 \kms interactions), and 
10$^{-7}$ (for a relative velocity of 100 \kms) under ISM densities of 2 \cm3. The temperature 
of the stellar wind is set at 10$^4$ \,K which is the lowest value for which 
the cooling function is defined in their simulation.

It is well known observationally that the mass-loss experienced by AGB stars is not constant,
and it has been shown that wind variations associated to the thermal pulses can lead to the 
formation of multiple shell structures, and large haloes around AGB stars (see e.g. 
\citealt{Vgm02,Vmg02,Sc05}). While realistic variable mass-loss rates along the AGB phase 
were explored in the study of the interaction process between an AGB star and the ISM 
\citep{Vgm03,Sze03,Vs05} we still lack a systematic study covering the range of expected relative velocities and ISM densities that
follows the evolution of the mass-loss from the star for different progenitors. 
This is precisely the goal of the present study. In this paper, we present simulations of
the formation of extended shells along the
AGB phase and their interaction with the ISM under a wide range of
conditions. In our models, mass-loss is not a free parameter, but follows
the stellar evolution prescriptions. This study is motivated
by the wealth of new observations of 
the resolved extended 
structure around AGB stars that are becoming readily available from  the {\it
  HERSCHEL Space Observatory} \citep{Lad10,Metal11,Jor11}, and will be complemented in the near
future with {\it ALMA}. 

The paper is organized as follows: in section \S2 we provide the
details of the numerical method, the initial and boundary
conditions, including a description of the parameters used to
characterize the environment, and the stellar dynamics; in section\S3 we describe the effects of the different ISM conditions during the early phase
of the evolution of the AGB when the stellar wind is constant; in \S4
we describe the simulations by showing the
evolution of the shells formed as the star ascends 
the AGB phase under different conditions;  \S5 we discuss the development
of instabilities in the shells; in \S6  the effect of
using different progenitor masses on the interaction; and in sections
\S7 and \S8 we present, respectively, a 
general discussion of the results and the conclusions of this work.

\section{THE NUMERICAL METHOD: INITIAL AND BOUNDARY CONDITIONS}

We have performed numerical simulations with the fluid
solver ZEUS-3D (\citealt{Sn92a,Sn92b,Smn92}), developed by
M.~L. Norman and the Laboratory for Computational Astrophysics. The
computations have been carried out on a
2D spherical polar grid with the angular coordinate ranging from 0 to
180\degree~and a physical radial extension of 4~{\rm pc}. The 
simulations have
resolutions of 800$\times$720 zones in the radial and angular coordinates of the grid
respectively, but a few models at lower resolutions (400$\times$360 radial and
angular coordinates respectively) have also been computed.
The models include the \cite{Rs:77} cooling curve above 10$^4$~{\rm K}. 
For temperatures below 10$^4$~{\rm K} the gas  is allowed to cool down with the 
radiative cooling curves given by \cite{Dm:72} and \cite{Mb81}.

Our boundary conditions are the AGB stellar wind and the parameters
that define the physics of the ISM. The evolution of the
star along the AGB phase is followed by feeding continuously the center of the grid with the  
stellar wind.  The mass-loss and wind temperature and velocity during the AGB phase has
been taken from \cite{Vw93}.  The wind temperature is assumed 
to be the effective temperature of the star. The simulations start at
the early-AGB phase, before the onset of the first thermal pulse, and continue 
until the end of the AGB phase. Further details of the wind assumptions and gas evolution
in a static configuration for different conditions can be found in \cite{Vgm02}.  

\begin{figure}
\plotone{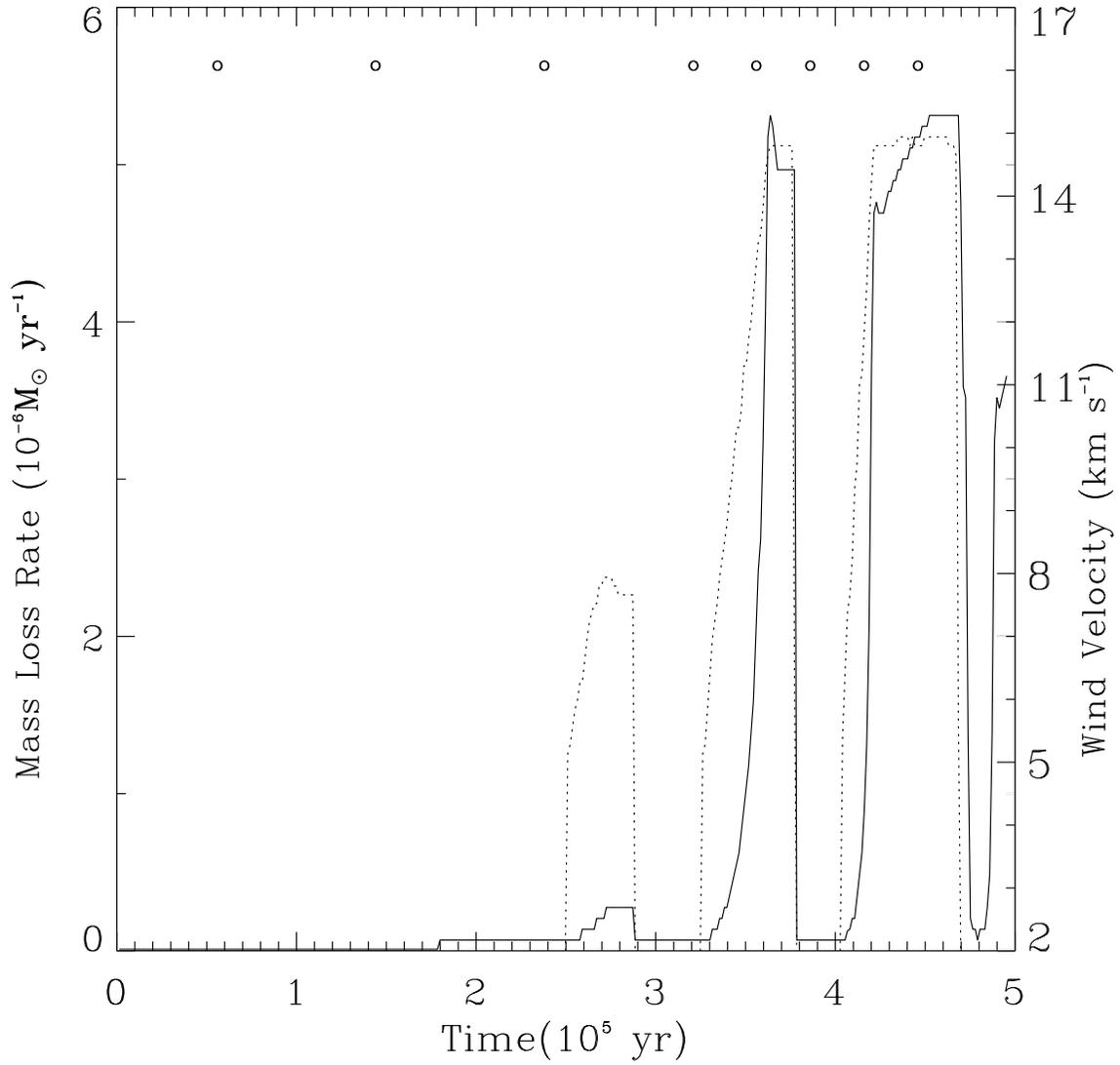}
\caption[ ]{Mass-loss rate 
    and wind expansion velocity used in
    the simulations of the 1 $M_{\odot}$ star along the AGB phase. 
    The solid line shows the mass-loss rate
    ($10^{-6}~M_{\odot}yr^{-1}$, left scale) and the dotted line
    shows the terminal wind velocity (\kms, right scale). The lowest mass-loss
    rates, not easily visible at the scale of the
    plot, are in the range 1-0.6$\times 10^{-8}~M_{\odot}yr^{-1}$ 
    with velocities between 2-5 \kms.  The circles in the top part of the
    plot mark the time at which the outputs of Fig.~4 are shown.
\label{agbwindevo1}}
\end{figure}

\begin{figure}
\plotone{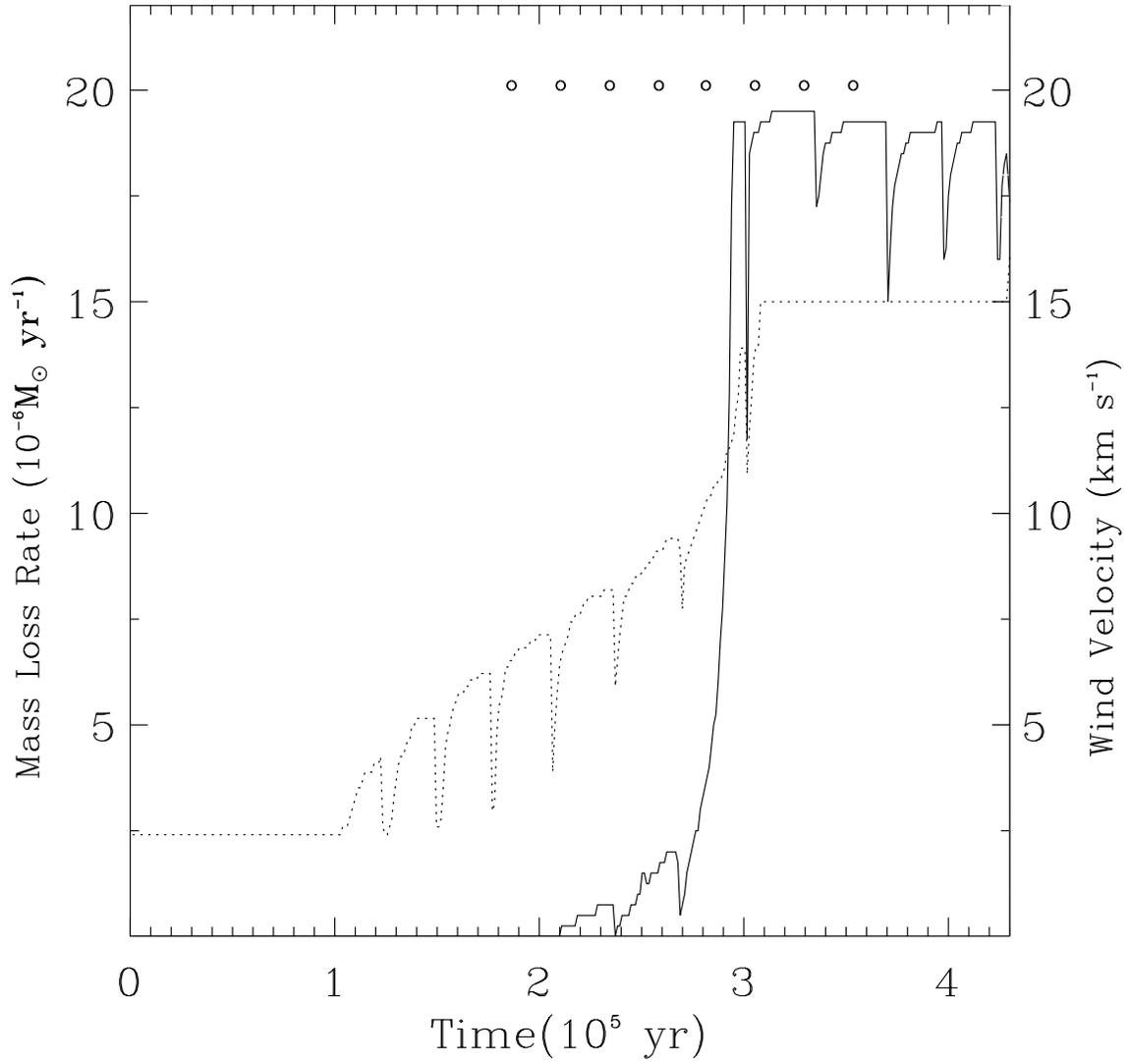}
\caption[ ]{Same as Fig.~\ref{agbwindevo1} but for a 3.5 \Mso star. 
     The circles in the top part of the
    plot mark the time at which the outputs in Figs.10 and 11 are shown.
\label{agbwindevo35}}
\end{figure}

To study the effect of the interaction on different 
progenitors, we have used stellar models for 1 and 3.5 \Mso~stars (Main Sequence 
masses). Fig.~\ref{agbwindevo1} shows the mass-loss (left axis, solid line) and
stellar wind velocity (right axis, dashed line) used as input for the
simulations of the 1\Mso~model during the AGB phase.  The same is
shown in Fig.~\ref{agbwindevo35} for a star with an initial mass of 3.5 \Mso.  

Finally, the interaction with the ISM is simulated by fixing the 
star at the center of the grid and allowing the ISM to flow into it at the outer
boundary from 0 to 90\degree. From 90 to 180\degree~we set an outflow
boundary condition. The temporal evolution 
of the stellar wind has been set within a small
(five radial zones) spherical region centered on the
symmetry axis, where reflecting boundary 
conditions are used. In doing that we have assumed that the ISM moves 
relative to the star perpendicular 
to the line of sight. Note that under the pure gas-dynamics scheme 
with linear artificial viscosity 
used here, shock errors are expected 
to be small \citep{Fal02}.

\subsection{The Environment: The Physical Conditions of the ISM}
Regarding the gas component of the ISM and using the temperature as a discriminator,
the diffuse ISM can be described within four major phases: the cold neutral medium
(up to 100~K), the warm neutral medium (hereafter WNM; with temperatures between 5\,000~K 
and 8000~K), the warm ionized medium (10\,000~K), and the hot ionized medium (10$^6$~K).  
The filling factors of each of the components
are still controversial \citep{Cox05}. The values proposed originally by
\cite{Mko} for the warm neutral (0.1), warm ionized (0.2) and hot ionized
component (0.7) have been revisited, mostly to account for a wide variety of
observations that point towards a much lower filling factor for the hot
component (less than $\sim$0.5).  The warm components are each thought to account
for $\sim$0.2 of the volume filling factor (see
e.g. \citealt{Cox05}). In short, most of the ISM volume relevant for this paper 
is occupied by hydrogen in its warm
and ionized forms (see e. g. \citealt{Bur88, Kk09}), and therefore,
for the simulations, most of the ISM has been chosen to have the
typical values of these two components. In this we have ignored the
solid ISM component to model the physical conditions of the ISM.  
Although the treatment of dust is fundamental when considering    
matter-radiation interaction processes in the ISM, its dynamical effects are
negligible given  its small mass with respect to hydrogen ($\sim 3\times10^{-3}$).  
We have not attempted to simulate the molecular gas component of the
ISM given that it is found mostly in the form of
discrete clouds occupying only a small fraction ($\sim$1--2 \%) of the interstellar 
volume (see e.g. \citealt{Fer01}).

The ISM densities used in this paper are within the range of values
observed for the warm ISM component from 0.1 to 1 \cm3
\citep{Kh88}. We have also explored typical values of the hot ISM 
component  (with densities below 0.003 \cm3
\citealt{Fer01}). We have only considered the thermal component
of the pressure to characterize the ISM, $P=n k T$ (where $k$ is the
Boltzmann's constant). Although all the phases of the ISM are thought to coexist in roughly
thermal pressure equilibrium, it has been shown that the average thermal
component is less than 1/3 of the total pressure in the mid-plane and that
the non-thermal pressure components (cosmic rays and magnetic fields) each
take roughly the other two thirds \citep{Fer01}.  Typical mid-plane values of
 thermal pressure are of
the order of $\sim 10^{-12}$ dyn cm$^{-2}$, decreasing outwards
\citep{Cox05}. 

The range of values used in our
simulations are between 10$^{-15}$ to
1.38$\times 10^{-13}$  dyn cm$^-2$ (with most of them
8$\times10^{-14}$  dyn cm$^{-2}$), therefore representing typical
values off the mid-plane (see Table~1). We have not included
magnetic fields or turbulent motions in the simulations.   

\subsection{The Relative Velocity: Stellar Dynamics}

To model the interaction process it is important to characterize dynamically the three
major stellar components of the Galaxy: the thin disk, the bulge, and the halo. Population
I stars belong to the disk and follow a differential rotation curve around
the center in nearly circular orbits with angular rotation rates a decreasing
function of their radial distance. Disk 
stars have a velocity  dispersion  10-40\kms \citep{Mb81}, which causes 
them to execute small oscillations about a perfectly circular orbit, both in
the Galactic plane (epicycles) and in the vertical direction. The
thin disk has a radius $\approx$ 25-30 Kpc and effective thickness $\approx$ 400-600 \, pc.  
We have characterized the population of stars belonging to the disk by using relative 
velocities in the 10--50 \kms~range (see Table.~1). Higher ISM densities 
have been used for the lower velocity models in order to
be consistent with the conditions found by stars moving closer to 
the mid-plane. For the same reason, we have decreased the ISM density when simulating stars moving with larger velocities away from the Galactic plane. Overall, the ISM density
range used spans two orders of magnitude (from 1 to 0.01 \cm3).   
 
How to characterize dynamically the stellar population of the halo is not so straightforward.
The halo extends out to more than 30 Kpc from the center 
\citep{Bin1} and the orbital behavior of the halo (population II) stars
is still not clear, but it seems to contain a combination of stars with
extreme retrograde orbits \citep{Car97}, highly inclined orbits, and stars
that do not move in regular orbits at all \citep{Els62}. In addition, there
are vertical and radial kinematic gradients (see e.g. \citealt{Maj93,
  Bee00,Cb00, Hel08}). To represent this population we have performed
simulations with larger relative velocities (85--100 \kms). 

We have not attempted to model the conditions relevant to the galactic bulge, 
although  any of the simulations described above can be  applied to
the lower end of the velocity range for AGB stars.

A summary of the parameters used for the simulations are given in
Table~1. Column (1) is 
the run ID name where the first number underscored refers to the stellar
mass; column (2) lists the velocity of the star relative to the
ISM; columns (3), and (4) give the value of the adopted ISM density and
temperature respectively; column
(5) lists the value of the ram pressure for the simulation and column (6) the
Mach number of the ISM; finally column (7) is the stand-off distance (see \S3).

\section{RESULTS: EARLY AGB CIRCUMSTELLAR SHELLS}

\subsection{The Early AGB Phase: Constant Wind}
During the early AGB phase, the evolution is characterized by a
constant free-streaming stellar wind with velocity $v_*\,_\mathrm{w}$
and mass-loss rate  
$\dot{M_*}\,_\mathrm{w}$. For a star
moving supersonically with a velocity  $v_\mathrm{ISM}$ through an ISM with
density $n_\mathrm{ISM}$, the distance from the star at which the ram
pressure of the 
free-streaming wind equals that of the interstellar medium is given  by \citep[see e.g.,][]{Vb90,Mac91}, 

\begin{equation}
r_\mathrm{so} =
5.5\times10^{17}\,\left(\frac{\dot{M_*}\,_\mathrm{w}}{10^{-8}}\right)^{1/2}\,\left(\frac{v_*\,_\mathrm{w}}{10^5}\right)^{1/2}\, 
\mu^{-1/2}\, n_\mathrm{ISM}^{-1/2}\, \left(\frac{v_\mathrm{ISM}}{10^5}\right)^{-1}~~,
\label{so}
\end{equation}
where $\mu$ is the dimensionless mean molecular weight, and
$r_\mathrm{so}$ is given in \,cm, $\dot{M_*}\,_\mathrm{w}$ in
$M_{\sun}~\mathrm{yr}^{-1}$, and the densities and velocities are given in \cm3
and $\mathrm{cm}~\mathrm{s}^{-1}$, respectively.  
 
Eq.~\ref{so} is applicable as long as the stellar wind is kept
constant at the inner boundary, or the wind is constant long enough to
reach pressure equilibrium with the ISM. For AGB stars that is the
case only during the early-AGB evolution (see Figs.~\ref{agbwindevo1} and
~\ref{agbwindevo35}). The validity of
this early stationary approach to calculate the analytical stand-off
distance is broken relatively early in the AGB evolution: at 
$\sim 1.8\times 10^5~\mathrm{yr}$ into the evolution of
the 1 \Mso~star, and at $\sim 1\times 10^5~\mathrm{yr}$  for the 3.5
\Mso~stellar model.  The values $r_\mathrm{so}$ calculated from Eq.~\ref{so} using
the early AGB wind (0.1$\times10^{-7}$\Mso and 2 to 2.5 \kms winds) 
are given in column (7) of Table~1.  

\begin{figure}
\plotone{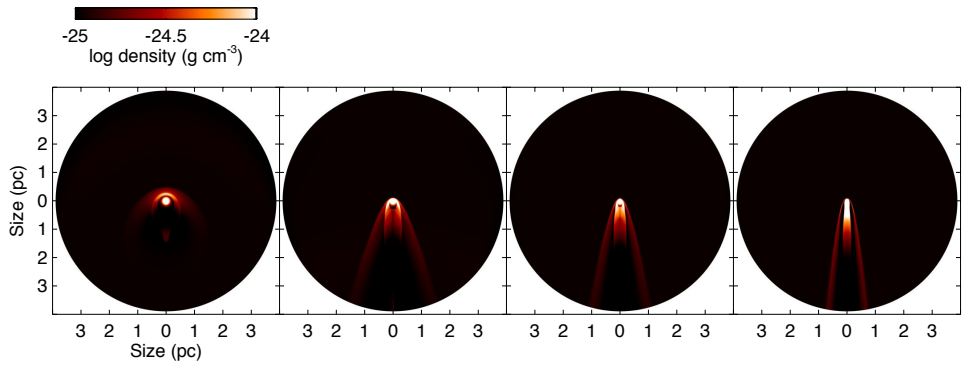}
\caption[ ]{The shell density is shown at 1.8
  $\times10^5\mathrm{yr}$ of the evolution of a 1 \Mso star in the
  AGB phase. The snapshots aim to illustrate the effect of increasing
  the relative velocity of the star with respect to the ISM and, therefore,
  they have been selected from simulations with the same ISM density
  equal to 0.1 \cm3. The relative velocities are 10, 30, 50 and 100
  \kms, increasing from left to right.  
\label{stationary}}
\end{figure}

At this early stage, the structure of the circumstellar shell in the simulations is
relatively simple, showing the characteristic bow-shock structure in
the direction of the movement. Fig.~\ref{stationary} shows the result
in the density structure, in logarithm scale, of the interaction
for relative velocities of 10, 30, 50, and 100 \kms (from left to
right). The panels have been taken at the end of
the stationary wind phase (at $\sim 1.8\times
10^5~\mathrm{yr}$) of the
evolution of a 1 \Mso~star through an ISM with a density of 0.1 \cm3.

The simulations show an asymmetric shell structure that
develops early in the evolution of the shell. Its morphology
is highly influenced by the relative velocity of the interaction (see
Fig.~\ref{stationary}). The most straightforward effect of increasing
the velocity of the interaction with the ISM is the closing of the opening angle of
the bow shock. It is important to note as well, that the size of the
cometary tail in the downward direction increases with the
relative velocity of the star with the respect to the ISM. The value of
$r_\mathrm{so}$, obtained from Eq.~\ref{so} and given in Table. 1, agrees well with the
sizes obtained from the simulations.  Note that comparing the
analytical and the numerical values is only meaningful for those models
in which the shell, at this stage, is larger than the grid resolution in
the radial direction (0.01 \,pc). 

It is important to note here that \cite{Wai06,War07,Wai07}, in their
study of the interaction with the ISM, 
only consider a single wind (with a constant mass-loss of either 
1, 5, 10, or 50 $\times 10^{ -7}$ \mlr depending on the velocity used for the interaction)
for the whole AGB evolution. A single AGB wind has been used as well by \cite{Retal08,Esq10}.
The only time in which the AGB wind is
constant in our simulations is during the early AGB phase described
above, and therefore, this is the only time at which it is meaningful to
compare the results of our simulations with the ones of the literature. 
A more detail comparison with the numerical simulations in the literature 
is given in the discussion section.

\section {RESULTS: THE EVOLUTION OF THE CSE AS THE STAR EVOLVES ALONG
  THE AGB PHASE} 
\subsection{Low velocity models: v = 10 and 20 \kms}
\begin{figure}
\plotone{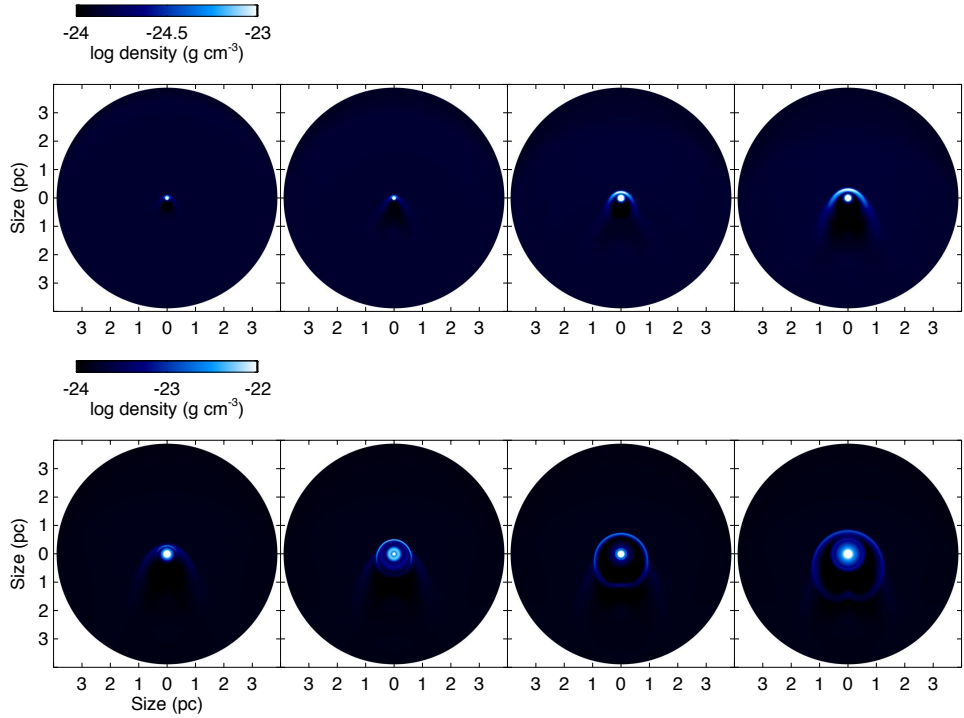}
\caption[ ]{Density (logaritmic scale) of the circumstellar envelope
  of a 1 \Mso AGB star moving at 10 $\mathrm{km}\,s^{-1}$. The ISM density
  is 1 ${\rm cm}^{-3}$. From left to right and top to bottom, the
  panels show the evolution of the shell at 0.6, 1.5, 2.4, 3.3, 3.6,
  3.9, 4.3, and 4.5 $\times10^5 \mathrm{yr}$ along the AGB. The times
  at which the models shown have been selected are marked in
  Fig.~\ref{agbwindevo1} where we plot the stellar wind used in the
  simulations. 
\label{v10n1}}
\end{figure}

Fig.~\ref{v10n1} illustrates
the structure of the Circumstellar Envelope (CSE) of a moving star as it 
evolves in the HR diagram along the AGB phase. Fig.~\ref{v10n1} corresponds
to the first model listed in
Table~1, Model R$_{1M}$~10h and aims to be representative of a population~{\rm
  I} star on the lower end of the expected velocity dispersion, 10\kms~
and evolving close to the Galactic plane. 

In Fig.~\ref{v10n1} from left to right and top to bottom, the  panels show the evolution of the shell at 0.6, 1.5, 2.4, 3.3, 3.6,
  3.9, 4.3, and 4.5 $\times10^5 \mathrm{yr}$ along the AGB.  In the first two
panels of Fig.~\ref{v10n1} the density structure is shown at a time
when the  stellar wind has still a very small momentum
 (10$^{-8}$ $M_{\sun}~\mathrm{yr}^{-1}$, 2\kms ). Yet the characteristic
feature of the
interaction has already developed: a bow shock in the direction of the
movement, that is, towards the top of the page. 

As the star evolves, (see Fig.~\ref{agbwindevo1})  the mass-loss 
rate (and wind velocity) increases and the CSE begins to grow in the
leading direction of the movement (third and fourth top panels of 
Fig.~\ref{v10n1}), given that the stellar wind has
enough momentum to compete with the ram pressure provided by the
ISM. In all, the stellar wind is always allowed to expand faster along
the opposite direction of the movement given the ambient pressure is
smaller and the opening angle of the bow-shock is maintained. It is 
important to note how different the 
shape of the CSE is along the evolution of the star on the AGB. This is
the result of the wind continuously changing in the inner
boundary of the grid. 

In the first panel on the bottom left, the star
is undergoing a second mass-loss increase, which is orders of magnitude larger
than any previous mass-loss. The stellar
wind propagates through the density profile created by the previously
ejected material and a shock region develops. This shock is not with
the ISM material but between subsequent episodes of mass-loss. No
shell is present  
in the opposite direction of the movement where
the wind expands subsonically and no discontinuity can develop.  It is
only towards the end of the AGB phase that the wind is dense enough to
create a shell in the direction opposite to the movement. 

Note that, in general, the stellar wind encounters a lower pressure opposite to the movement
and is able to expand further in that direction. As a consequence the star is
displaced from the geometrical center of the shell and an outer
asymmetric CSE is formed.

The ISM density used in this model, 1 \cm3, aims to represent the average
conditions of the cold neutral medium in the galactic plane, and
accordingly the temperature of the ISM gas was set at 100
$\mathrm{K}$. The large Mach number of the ISM, in this case results in a
larger shock compression given that the shell is cooling radiatively. 

It is expected that a large number of AGB stars might  be moving
at such low velocities relative to the local ISM. A large ISM density, 
even though associated to a  
very slow moving star, can generate an asymmetric 
shell. The stellar wind gets
deflected by the high ram pressure of the ISM and expands more easily
in the direction opposite to the interaction. The stellar wind always
expands inside the cavity cleared by the previous stellar activity,  
not showing any sign of the interaction process. The asymmetry of the interaction is only
observable in the outermost shell. For the parameters used in this model a large 
asymmetric outer shell is formed which is a factor of two
larger in the downstream direction compared to the upstream direction.  
No morphological feature of the interaction is recorded in the
inner CSE at the end of the AGB given that the wind is expanding
almost unperturbed within the asymmetric outer cavity created by the interaction.
No tail in the opposite direction of the movement remains in this
model. The bulk of the mass lost by the star is left behind along the
movement, only the latest 10\,000 $\mathrm{yr}$ of mass-loss  are recorded
in the shell structure.

\begin{figure}
\plotone{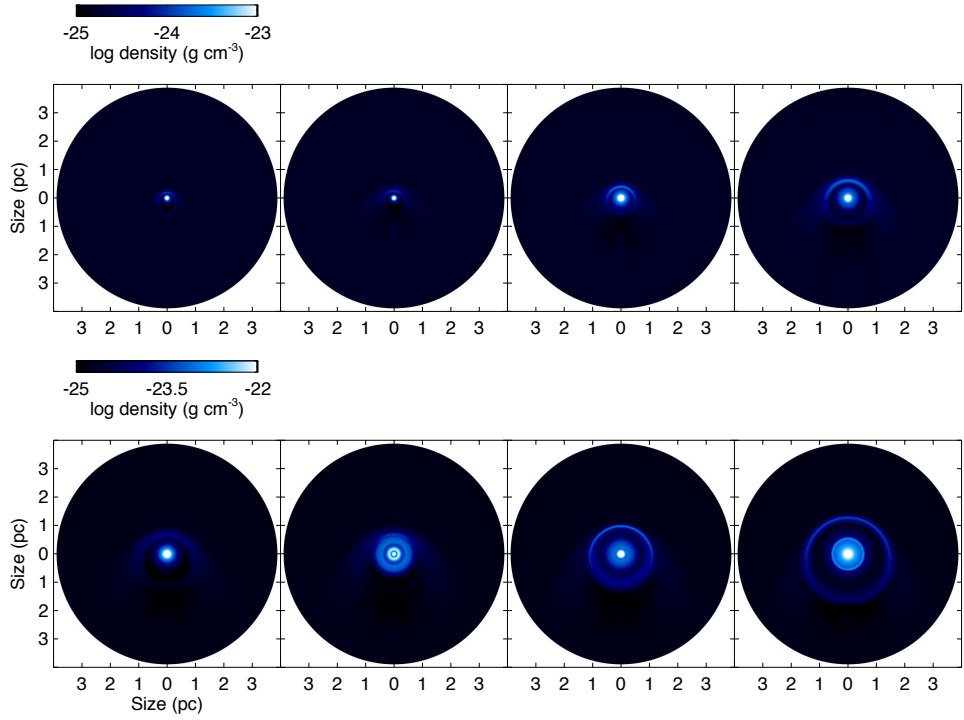}
\caption[ ]{Same as Fig.~\ref{v10n1} but for an ISM with density 0.1 ${\rm cm}^{-3}$.
\label{v10n0.1}}
\end{figure}

Fig.~\ref{v10n0.1},  model R$_{1M}$~10l, shows the CSE density structure 
formed by the
interaction of a 1 \Mso  AGB star with an ISM, with a density  10
times lower than the
one shown in Fig.~\ref{v10n1}. 
The snapshots in the Figures have always been chosen
at the same time in the evolution, unless noted otherwise, in order to better
show the effects of changing the conditions of the environment.

The main morphological features of the interaction
described previously are also present here and, as in the previous case, the mass lost by the
star is continuously deflected in the downstream direction of the
movement. The important difference in this case is that a smaller ISM
density allows for the formation of a shell in the downstream
direction and the lower ram pressure of the ISM allows a larger growth of the
shell in the upstream direction. As a consequence
the asymmetry generated by the interaction
on the outermost shell is less obvious in this case, and translates to a
small displacement of the location of the central star with respect to
the geometrical center of the envelope once the mass-loss rate
associated to the last thermal pulses takes place (last two bottom
panels). In general, once the mass-loss reaches its highest rate at the end of the
AGB, the effect of the interaction on the shell is very small. 

Models evolving at relative velocities of 20 \kms at densities of 0.1
and 0.01 \cm3 were presented in \cite{Vgm03}. Note that in model R$_{1M}$~20l 
we have used an ISM density of 0.01 \cm3 resulting in the lowest ram pressure 
considered in this work. This models show the 
same main features of the models described above. The AGB
wind forms a bow shock upstream of the central star. The mass lost by the star as it
ascends the AGB does not interact directly with the ISM, it expands
within the bow shock and gets deflected downstream. It is only the
mass lost at the end of the AGB
phase that generates enough pressure to reach the bow-shock. This
mass-loss also leads to the growth of the bow-shock structure in the
direction of the movement and also generates shocks with the stellar mass-loss 
within this shell. The
asymmetry is still noticeable at the end of the AGB, the bulk of
the mass lost by the star is deflected in the downward
direction. Besides losing the spherical symmetry, the CSE  formed
under a low velocity interaction has a smaller size and 
contains a lower mass than when formed under zero relative velocity \citep{Vgm02}.
This statement 
states true also for the models moving at 10 \kms.

\subsection{Intermediate Velocity Models: 30 and 50 \kms}

\begin{figure}
\plotone{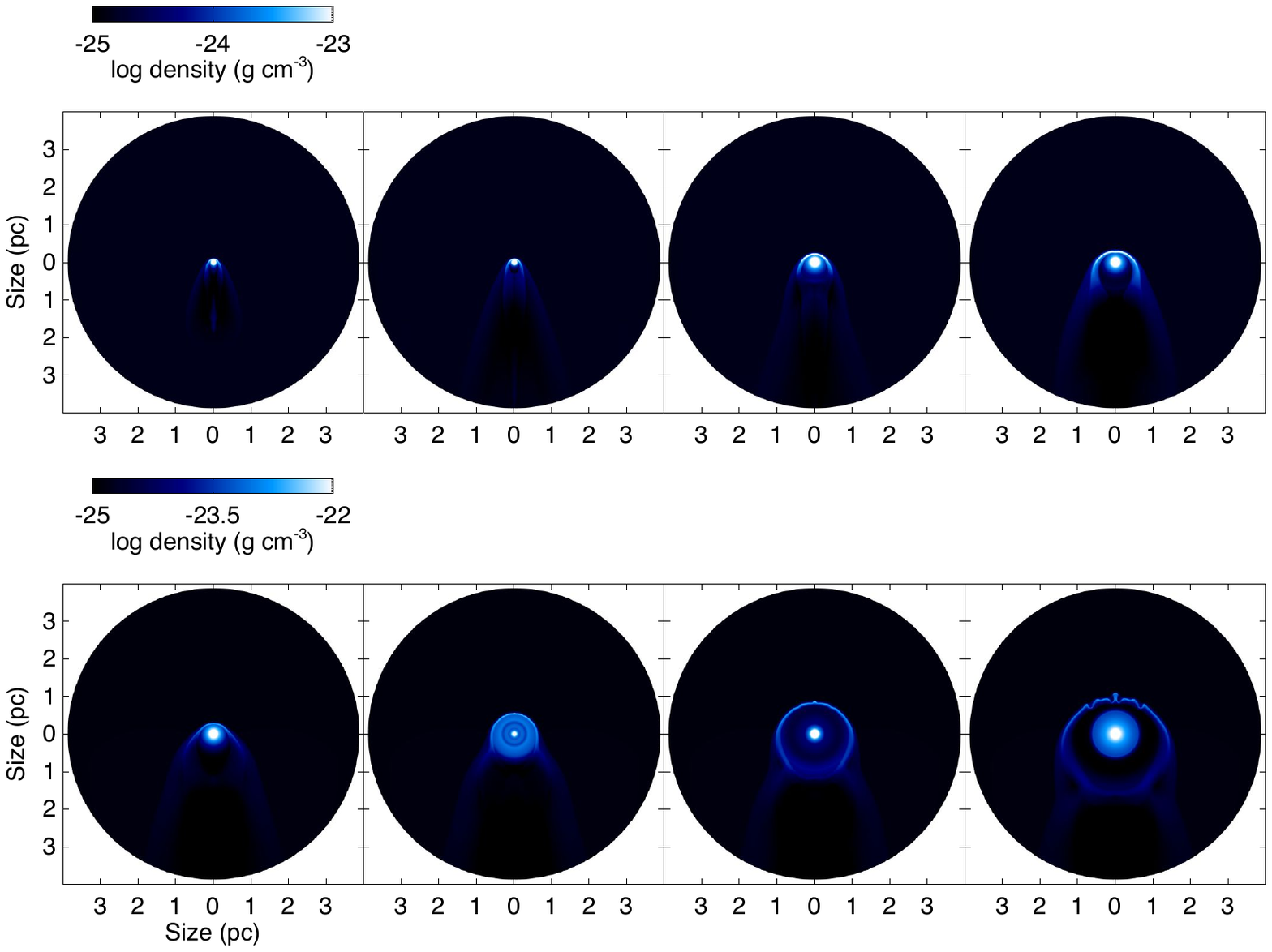}
\caption[ ]{Same as Fig.~\ref{v10n1} but for a star moving at 30
  $\mathrm{km}\,s^{-1}$ through an ISM with density 0.1 ${\rm cm}^{-3}$ 
\label{v30n0.1}}
\end{figure}

\begin{figure}
\plotone{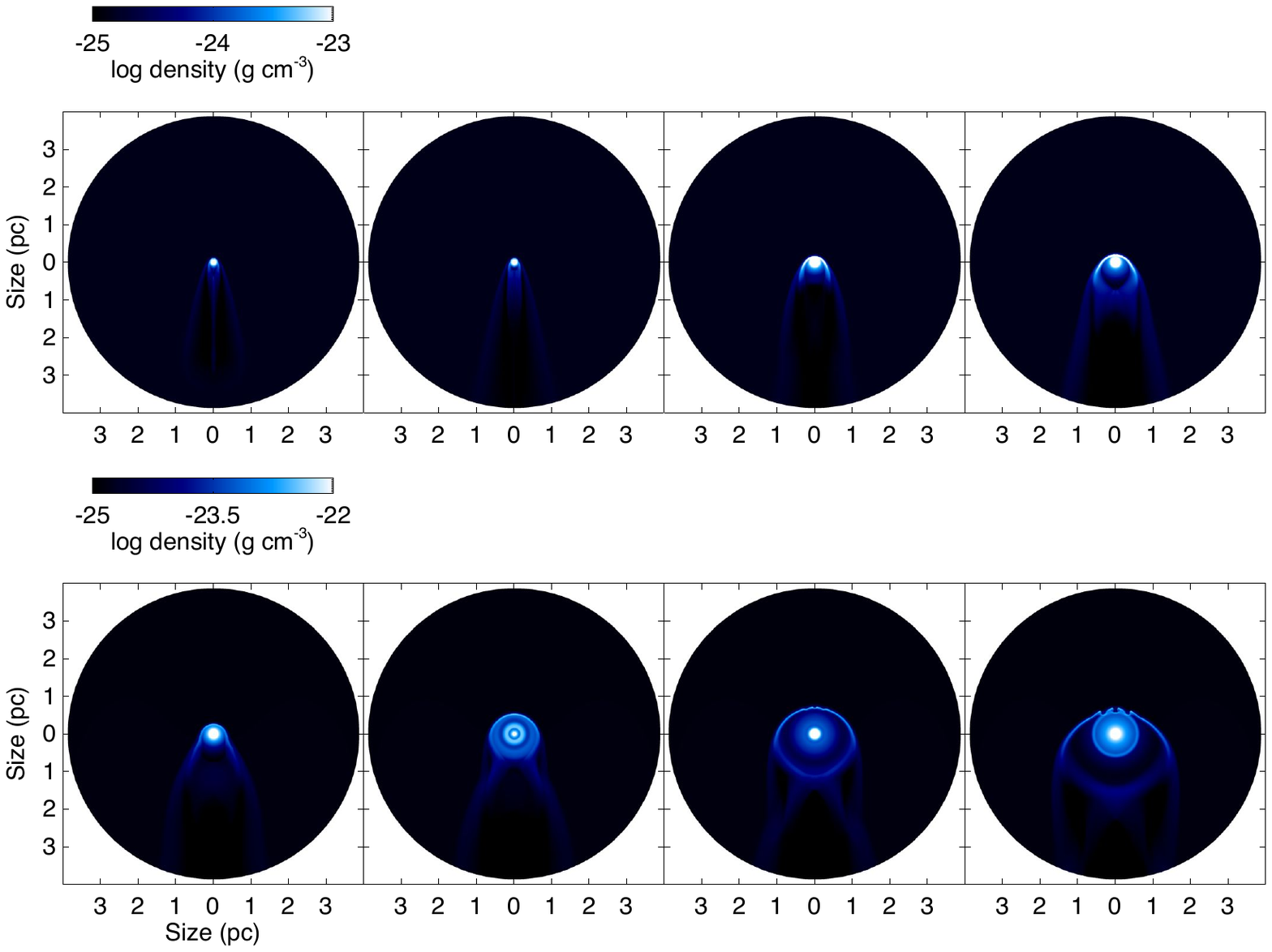}
\caption[ ]{Same as Fig.~\ref{v10n1} but for a star moving at 50
  $\mathrm{km}\,s^{-1}$ through an ISM with density 0.1 ${\rm cm}^{-3}$ 
\label{v50n0.1}}
\end{figure}

Figs.~\ref{v30n0.1}, \ref{v50n0.1} show simulations of a low-mass star
moving with intermediate relative
velocities, 30 and 50\kms~, through an ISM with a density of
0.1 \cm3. These models aim to match the velocity dispersion of stars belonging to the disk
population. As with the figures
shown in  the previous subsection, the sequence of outputs have
been selected at the same times as shown in Fig.~\ref{v10n0.1} and
marked in the top  of Fig.~\ref{agbwindevo1}.

As for the low velocity models shown before, in the first two top panels the AGB
wind still has the characteristics of the early AGB phase, a bow-shock
forms in the leading direction. The stationary wind phase
lasts long enough for the AGB wind to reach pressure equilibrium with
the ISM. Thus the bow-shock reaches a stable position ahead of the
star. Both the opening angle of the bow-shock, and the distance from
the star, decrease as the velocity that characterizes the ISM interaction increases. 

The main difference between the models with low and intermediate
velocities can be seen already in the third and fourth
top panels of Figs.~\ref{v30n0.1} and \ref{v50n0.1} shown at 2.5$\times10^5$ and 3.4$\times10^5 
\mathrm{yr}$ into the AGB evolution. For these intermediate relative velocities the 
previous mass-loss gets deflected behind the star more efficiently and thus the stellar 
wind encounters the ISM directly. The main characteristic of the interaction 
for these models is the development of instabilities in the bow-shock.  
When the wind expands throughout surroundings already cleared by
the previous stellar activity, no instabilities are developed. That is the case for the
models with velocities of 10 and 20 \kms. However, as the velocity
of the interaction increases, the outermost shell is subject to instabilities 
as predicted by \cite{Bk98} (see \S6). 

The wind expands faster along the hot cavity left behind by the star. The formation of a
highly asymmetric shell that grows in size in the perpendicular and in the
opposite directions of the movement  can be seen in the shell evolution 
in Figs.~\ref{v30n0.1} ,\ref{v50n0.1}. Again the leading part of the CSE
expands away from the star only when the stellar wind has enough momentum (either
the wind velocity increases or the mass-loss rate) to
compete with the ISM ram pressure. And, in the few
instances that the ISM pressure is larger than the one provided by the
shell in the leading direction, the ISM pushes the shell inwards
creating the seeds for the formation of an unstable flow. The
formation of instabilities highly influence the CSE shell structures as the unshocked ISM is
able to penetrate deep into the shell as it breaks up. 

In the downstream 
direction the wind expands 
within the tunnel left behind by the star and an elongated tail grows.
Mass is constantly flowing away from the head of the bow
shock, feeding the cometary structure. The tail material
is formed by a mixture of material stripped from the head of the bow
shock (and cooling as it flows downstream around the bow shock
towards the back of the star) and wind material directly ejected  by
the star in the downstream direction.

Decreasing the density of the ISM for the same velocity has an
enormous influence in the shell evolution as shown in model R$_{1M}$~50l.
In fact, this model has a ram pressure comparable with the first  model listed
in Table.~1 for a velocity of 20 \kms and ISM density of 0.1 \cm3 
 \citep{Vgm03}. Both models show that the effects of the
interaction in the morphology of the outer shell are small
towards the end of the AGB evolution. However, the tail in the opposite
direction of the movement is still formed as ram pressure is stripping
matter from the shell. Evolution through low ram pressure environments
($\approx 10^{-13}$ $\mathrm{dyn}\,\mathrm{cm}^{-2}$) does not produce 
a strong morphological feature at the tip of the AGB. However, no
matter how small the ram pressure of the interaction is, the
morphology is strongly affected as the star ascends the AGB. Still this
effect is important in the
evolution of the CSE shells. Ram pressure stripping constantly removes
the mass of the circumstellar envelope, and as a consequence lower density
shells are formed. The external pressure provided by the external
medium reduces the expansion velocity of the shell in the direction of
the interaction, increasing it along the opposite one.  

\subsection{High velocity models: 85 and 100 \kms}
\begin{figure}
\plotone{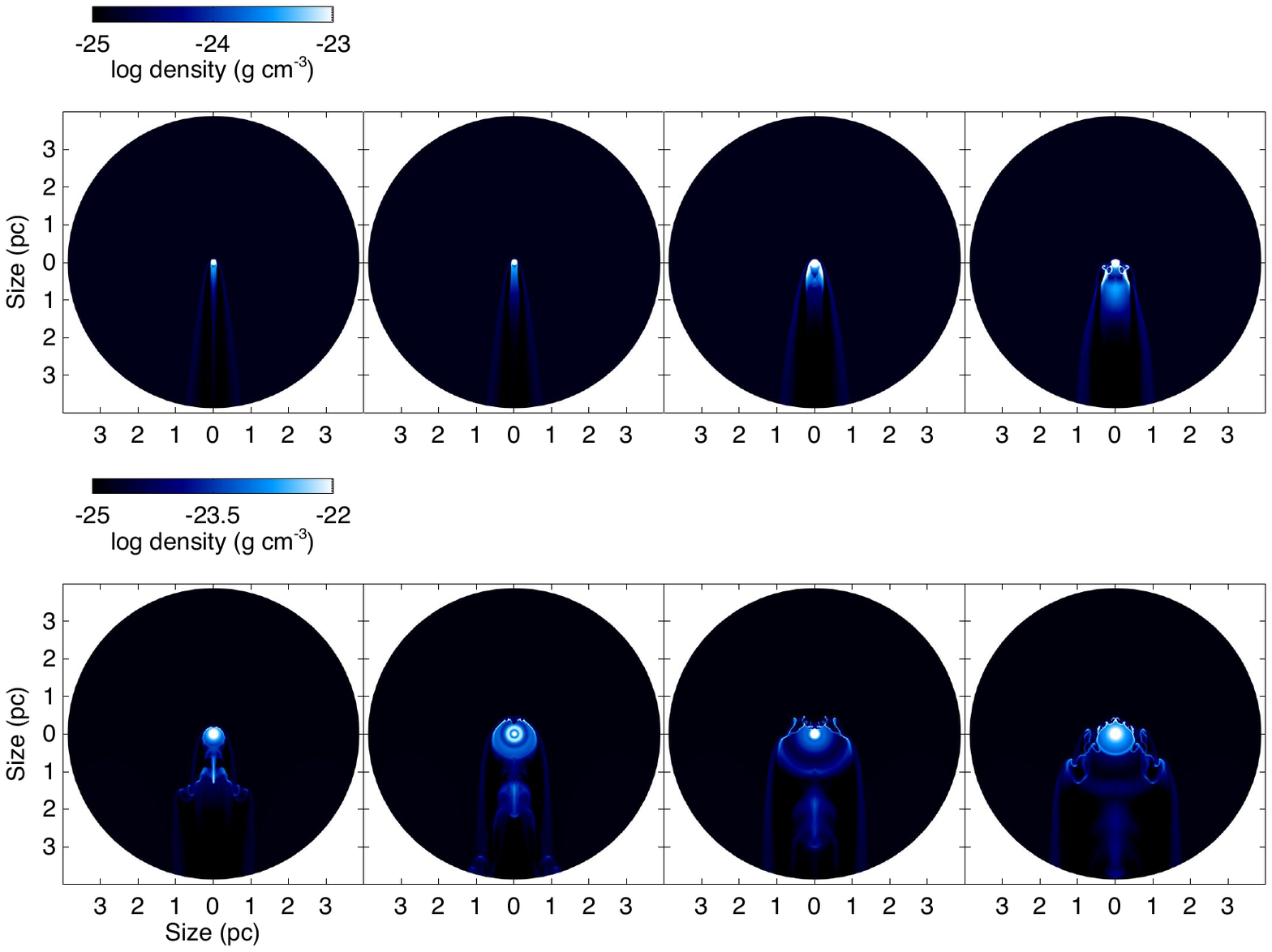}
\caption[ ]{Same as Fig.~\ref{v10n0.1} but for a star moving at 100
  $\rm km\,s^{-1}$ through an ISM with density 0.1 ${\rm cm}^{-3}$.
\label{v100n0.1}}
\end{figure}

A relative velocity of  100 \kms for the ISM interaction is shown in
Fig.~\ref{v100n0.1}, model R$_{1M}$~100h.  A narrow, confined bow shock is formed in the
upstream direction and a long tail is prominent downstream of the
motion. This model has the largest ram pressure of all the models
computed and shows the strongest features of the interaction.
At the time when the first increase in mass loss takes place (third
top panel from the left) the flow
is already dynamically unstable, and soon the bow shock is broken by
the instabilities  (see fourth top panel). The tail formed in the 
downstream direction is Kelvin-Helmholz
unstable. In the tail wind material expanding outwards from the
star interacts with
material generated in the eddies of the outer shell, material that is
moving inwards towards the star. A turbulent region develops  where
they encounter each other and several shocks are formed within the tail.

The stellar wind interacts directly with the ISM, it
becomes unstable, breaks and gets deflected. The ISM can
penetrate in the leading direction close to the star when
the wind has its minimum mass-loss between superwind events.  At the time of the last
increase in the stellar mass loss, the shell in the leading
direction grows in size despite being highly unstable. At the end of
the AGB the CSE shows a largely turbulent  morphology with several 
condensations caused by the shell fragmentation. The overall morphology
of the bow shock is maintained by tracing the tips of the condensations
caused by the instabilities. A similar unstable flow has been shown to develop in model  
R$_{1M}$~85l (v=85\kms n= 0.05\cm3) \citep{Sze03}.  

Fig.~\ref{v100n0.01} shows the same as Fig.~\ref{v100n0.1}  but
reducing by a factor of 10 the density of the ISM, which can be consider 
more realistic given the high
velocity considered. A lower density of the interaction reduces the
ram pressure and allows us as well, to study the effect of the ISM density 
in the interaction. In this model, R$_{1M}$~100l, the flow is dynamically 
unstable as well. However, given the lower
ram pressure the ISM cannot penetrate close to the star when the
stellar wind reaches its minimum. So the mixing between ISM and wind
material is not so efficient. The tail in the downstream direction is very
prominent in this model given that the ISM density is lower and
therefore any density enhancement has more contrast. The material gets
deflected downstream and mixed efficiently on the tail of the
interaction while in the direction of the movement the shell, although
unstable, does not break up completely and in fact it is allowed to
grow in the direction of the movement. 

Note that this model R$_{1M}$~100l (v = 100\kms n = 0.01\cm3) has the
same ISM ram pressure as model R$_{1M}$~10h (v=10\kms,  n=1 \cm3). 
However, in order to adopt realistic ISM conditions the temperature of the ISM
had to be modified according to the ISM density. As a result, the Mach
number of the ISM is very different in these three cases (see Table 1).   
Despite the fact that these models have the same stellar input and that
they are evolving against the
same ram pressure of the external medium the CSE formed are radically
different. The R$_{1M}$~10h  model does not develop instabilities, 
the outer shell does not break up, and a cometary tail
does not form either. This shows that ram pressure is
not the main parameter in order to determine the shell morphology. 
Ram pressure, however determines the sizes of the outermost shell 
at the end of the AGB evolution, and it can be seen that the shell sizes are 
the same for the two models (when tracing the tips of the condensations
caused by the instabilities).

\begin{figure}
\plotone{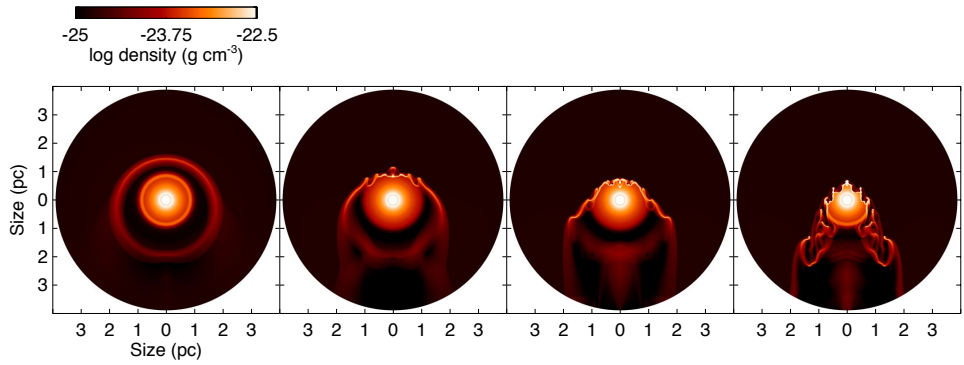}
\caption[ ]{Snapshots taken at 4.74$\times10^5\mathrm{yr}$ of a 1\Mso
  star moving (from left to right) at 10, 
30, 50, 100 $\rm km\,s^{-1}$.
\label{combinav}}
\end{figure}

In Fig.~\ref{combinav} we show the same as in Fig.~\ref{stationary}, but
with the outputs taken later on in the evolution. All the models
evolved through the same ISM density of 0.1 \cm3 and the relative
velocities, from left to right, are 10, 30, 50, and 100 \kms. 
Before we pointed out that  the most straightforward effect of increasing
the velocity of the interaction with the ISM is the closing of the opening angle of
the bow shock. Here we need to emphasize the more prominent effect of
the instabilities in the shell morphology as the relative velocity of
the interaction increases. It is important to note as well, that the
cometary tail in the downward direction becomes more prominent as the
relative velocity of the star with respect to the ISM increases.   

\section{MIRA AS A TEST OF A HIGH VELOCITY INTERACTION}
 
Mira is a thermally pulsating AGB star that is expected to experience wind modulations 
associated with its thermal pulse cycle. CO Observations of the system trace the current and/or 
most recent mass-loss to be 1.7$\approx\times10^{-7}$ \mlr 
\citep{Ry01}. Mira is moving at $\approx$ 125 \kms through its surrounding medium. Located at 
107 \,pc \citep{Kn03} the physical size of its far-ultraviolet cometary tail and bow-shock 
head are 4 and 0.1 \,pc respectively. The tail has also been detected in HI out to 0.4 pc 
from the star \citep{Mat08}. Herschel's PACS and Spitzer observations reveal that the arcs 
seen around Mira's head likely result from a combination of the projected 3D 
structures resulting from the interaction of Mira's wind with its companion on one hand, 
and with the ISM on the other hand \citep{U08,Mey11}. 

Although we have not run any specific model to match the observations of 
Mira \citep{Metal07}, we have a number of simulations with a range of parameters similar to those
used in the literature specifically to devoted to that matter \citep{Wai07,Esq10}, that allow us 
to perform a limited qualitative comparison among models. We are excluding from the comparison: 
i) the models of \cite{Retal08} since they focused on reproducing the complex double 
bow-shock in the cometary head using a latitude dependent wind associated with the binary 
component, and ii) the models of \cite{Rc08} given their analytical nature. 

In Table~2 we have summarized the relevant parameters to allow the comparison: column (1) 
gives the relevant reference (where models R$_\mathrm{1M}$~100l and R$_\mathrm{1M}$~100l allude to
this work); column (2) gives the relative velocity of the interaction; columns (3) and (4) give 
the density and temperature adopted for the ISM; finally columns (5), (6), and (7) provide 
the stellar wind parameters adopted in the simulations in terms of mass-loss, wind velocity and 
temperature respectively. 

\begin{figure}
\plotone{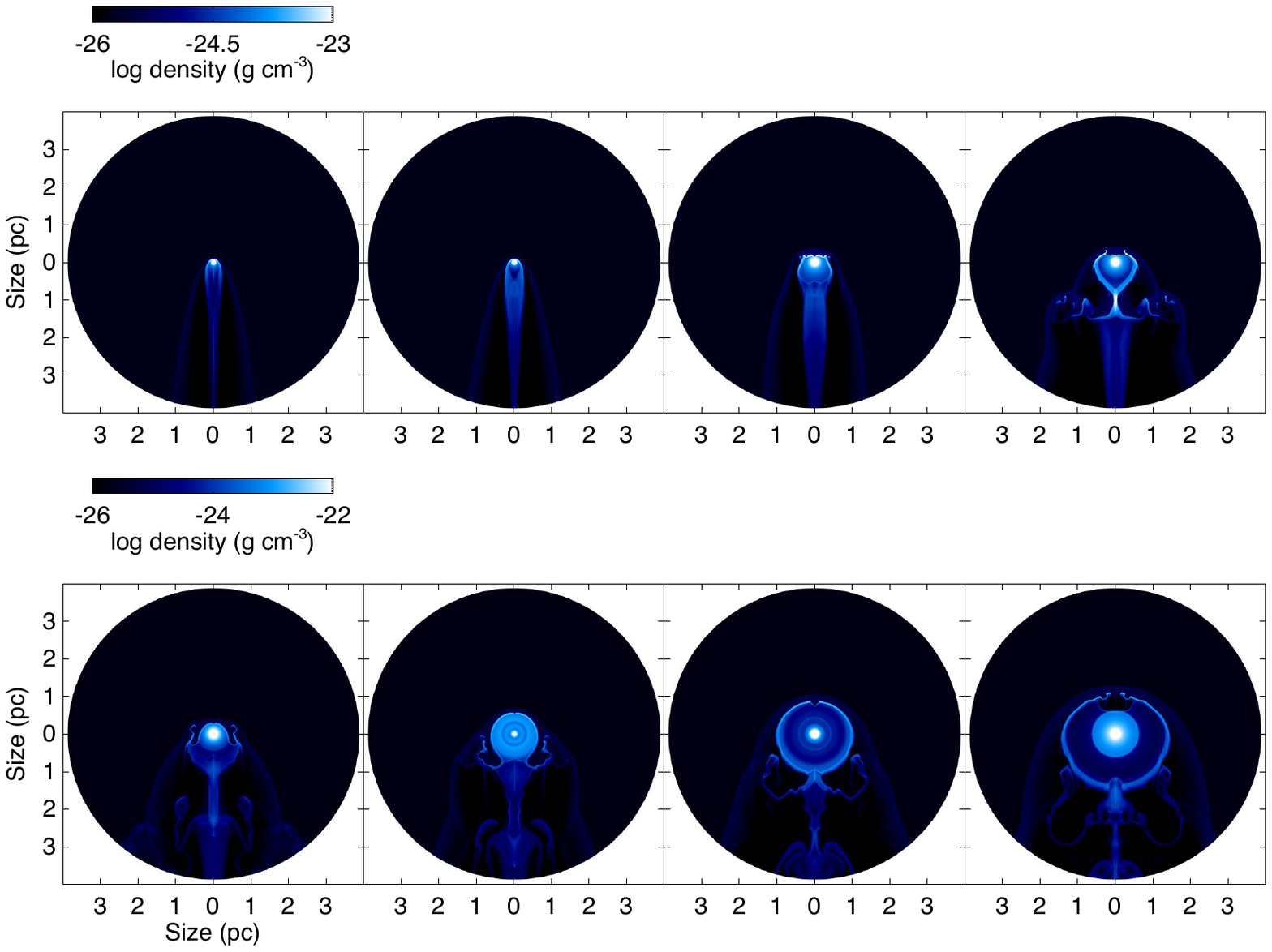}
\caption[ ]{Same as Fig.~\ref{v10n0.1} but for a star moving at 100
  $\rm km\,s^{-1}$ through an ISM with density 0.01 ${\rm cm}^{-3}$.
\label{v100n0.01}}
\end{figure}

Note that in our models, the wind is always changing in the inner 
boundary according to what is expected from a thermally pulsating AGB star, however 
until  $\approx 1.8\times10^5$ \,yr into the evolution the wind is constant, and until 
$\approx 2.8\times10^5$ \,yr 
although variable in mass-loss, wind velocity and temperature still has parameters similar to
those used in other models to compare with the observations of Mira. The outputs of the simulations that are relevant to this discussion are those shown in the top four panels of 
Figs~\ref{v100n0.1} and \ref{v100n0.01}. Note that only the first two ones are computed under 
a constant wind assumption.

As in previous models, the location of the stand-off distance and the size of the tail are 
similar to those observed once we take into account the slight differences in both the wind 
parameters and the ISM conditions for the interaction (see first panel in the top left of
Figs~.\ref{v100n0.1} and \ref{v100n0.01}). In both of our simulations, under a low (0.01 \cm3) 
and high (0.1 \cm3) density environment we obtain narrow cometary tails as a consequence of the 
interaction. Note that although the model evolving in a low density environment is similar to 
that computed by \cite{Wai07}, we obtain, however, a very different cometary structure. 

The model of \cite{Wai07} 
failed to reproduce the broad-head narrow tail structure observations of Mira that \cite{Esq10}
matched by changing the conditions of the environment (i.e. a dense ISM is changed into a low 
density rarefied medium with the parameters of the local bubble). The reason for this behavior
must lie in the unrealistic temperature for the AGB wind used by \cite{Wai07} (10\,000 \,K) and 
the fact that the gas is not allowed to cool down below this value in their simulations. If we 
assume that the ISM temperature is the same as that used in other works by the same authors 
(the value is not given in \citealt{Wai07}) then the AGB wind has always a larger temperature 
than the ISM. The cooling function, and the temperature assumed for the wind have an important 
effect in the formation of the tail. Higher density regions formed behind the star will 
cool more efficiently and will collapse against the ISM pressure allowing the formation of 
narrow tails as seen in our simulations. Narrow cometary tails are also formed in
the simulations of \cite{Esq10}, but they require a change from a high to a low density 
environment. A more direct quantitative comparison between our models and those of \cite{Esq10} 
is not possible given the too different conditions assumed. 

Although a change in the physical conditions of the ISM is possible, according to our models 
it is not necessary to explain the overall morphology of the Mira shell. A narrow tail is formed 
under standard evolution early in the evolution of the star. 

\section{DEVELOPMENT OF INSTABILITIES AND FRAGMENTATION}
Same of the computed AGB circumstellar shells are heavily fragmented,
see e. g. models R$_\mathrm{1M}$~30h, R$_\mathrm{1M}$~50h,
R$_\mathrm{1M}$~100h (also R$_\mathrm{1M}$~85 in \citealt{Sze03}). This strong fragmentation is 
caused by several processes.  The main cause of instabilities, as discussed extensively 
by \cite{Bk98} in their high resolution calculations
of 2D and 3D isothermal stellar wind bow shocks, is
the Nonlinear Thin Shell Instability (NTSI) \citep{Vis94}, and 
the Transverse Acceleration Instability (TAI) \citep{Dgani} when the bow shock is 
radiative (nearly isothermal) both in the forward shock as well as in the reverse shock 
of the stellar wind. This instability depends primarily on the Mach number 
of the star moving through the ISM,
requiring a minimum Mach number of a few. The effect of the instability is to ripple the
bow shock with wavelengths and amplitudes of the order of the nominal standoff distance 
of the bowshock. Based on the internal dynamics of the slab, the overall evolution 
of bends in the slab, and the fact that the instability grows faster near the apex 
of the bowshock, \cite{Bk98} argued that the instability of isothermal 
stellar wind bowshocks could be attributed to the NTSI. The NTSI is driven by shear 
flows within the shell created by large-wavelength wiggles in the shell.
Once the large distortions created by the NTSI advect into the wings of the bowshock, the action
of the TAI becomes important in further distorting the shape of the bowshock (see Fig.~4 from
\citealt{Bk98}). We believe that this is the case for models R$_\mathrm{1M}$~30h, 
R$_\mathrm{1M}$~50h, R$_\mathrm{1M}$~100h.

It has also been argued that Red Super Giant (RSG)  (similar to AGB in mass loss rates 
and wind velocities)  bow shocks are subject to 
Rayleight-Taylor Instabilities (R-TI) \citep{Be95a,Be95b,vanetal11}, since
the RSG, or AGB shocked gas, and the contact discontinuity are decelerated by the shocked ISM.
We think that this is the case for our model  R$_\mathrm{1M}$~100l, as we discuss below, 
and not the NTSI as also discussed by \cite{Bk98} for the case where the forward shock 
is adiabatic (in our model R$_\mathrm{1M}$~100l, the forward shock is semi-radiative).

\cite{Wai07c} have also argued that the main source of instability in AGB wind bowshocks 
is due to the  transverse shear of the flow, or the Kelvin-Helmholtz Instability (K-HI). 
However, as it has been discussed by \cite{Bk98}, this cannot be the case, the main reason 
is that the shear is only important in the wings of the bowshock. Although the resolution used
by \cite{Wai07c} was no sufficient to claim any quantitative analysis of the
instability, they argued that the K-HI  was the reason for the formation of their computed vortex.
Note however that their vortex starts near the apex of the bowshock 
(panels A, B, C and D in their Figure 2), where there is no transverse shear as pointed out 
by \cite{Bk98}.  Note also that the shape (spiral)  of the vortex is against
the motion of the flow, contrary to the vortex expected by the K-HI, where
the spirals form in the direction of motion of the flow. 
Even more, panels C and D in their Fig.~2 show an 
oblique shock (the vortex) piling up gas against the contact discontinuity, 
which is the typical behavior of the NTSI. 

We then have two different scenarios in our calculations, according to the amount of
radiative cooling in the forward shock. In the first one, the forward shock is radiative
(similar to the isothermal case), while in the second one, the forward shock is semi-radiative.
To show the difference we can analyze models R$_\mathrm{1M}$~100l and R$_\mathrm{1M}$~100h, 
both with the same Mach number respect to the motion in the ISM. 
The forward shock in the case
of R$_\mathrm{1M}$~100l is smooth (last panel in Figure 9), and is relatively 
well separated from the contact 
discontinuity. 
In this case, the fragmentation which starts in the apex is purely due to R-TI.
However, the forward shock in the case R$_\mathrm{1M}$~100h (last panel of Figure 8) 
is corrugated and its
location is quite close to the contact discontinuity. This is due to a stronger radiative cooling in
the ISM shocked gas. In both cases, the involved velocities are small, so we can assume 
that the bow shocks are in the radiative regime (the post-shocked gas temperature is below
$10^5$ K). The theory says that in these cases, the post-shocked gas density should be
the square of the Mach number multiplied by the ISM density (see also \citealt{Bk98}).
Thus, the shocked ISM gas density in model R$_\mathrm{1M}$~100h would be 12.1, 
while in the case of model R$_\mathrm{1M}$~100l would be
1.21 (see Table 1). Since the radiative cooling depends on the density
as $n^2$, the difference in both cases is 
notorious.  Thus, the  Mach number is not the only important parameter as discussed by
\cite{Bk98}, the density of the ISM is also important.
We can conclude that the case R$_\mathrm{1M}$~100l is "semi-radiative" if compared with 
R$_\mathrm{1M}$~100h (the cooling here is one hundred times larger). 
Thus, when the radiative cooling is strong enough (like the case of R$_\mathrm{1M}$~100h), 
the forward
shock is located very close to the contact discontinuity, and it is influenced by its shape, which
in this case is unstable to R-TI. Once
that the forward shock is corrugated, the oblique shocks of the NTSI exacerbate the 
fragmentation of the AGB circumstellar shells. 

In the cases computed here, the development of the fragmentation either by R-TI or by NTSI
is exacerbated during the inter-pulse periods, since the ram pressure of the AGB wind
drops considerably compared to the dynamical time.

\section{STELLAR PROGENITOR MASSES}

\begin{figure}
\plotone{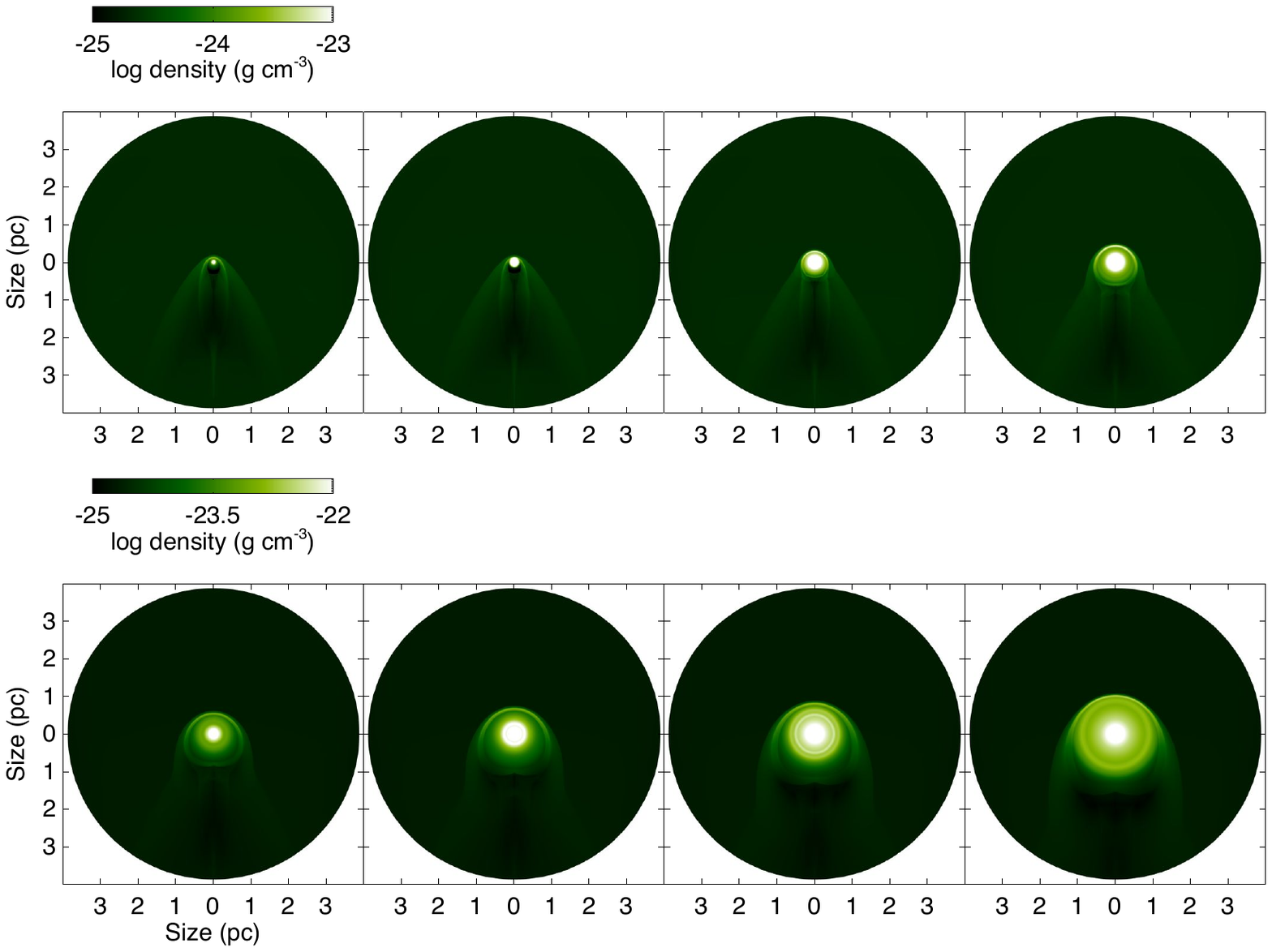}
\caption[ ]{Density (logaritmic scale) of the circumstellar envelope
  of a 3.5\Mso AGB star moving at 20 $\mathrm{km}\,s^{-1}$. The ISM density
  is 0.1 ${\rm cm}^{-3}$. From left to right and top to bottom, the
  panels show the evolution of the shell at 1.9, 2.14, 2.4, 2.6, 2.9,
  3.1, 3.3, and 3.6 $\times10^5 \mathrm{yr}$ along the AGB. The times
  at which the models have been selected are marked in
  Fig.~\ref{agbwindevo35} where we plot the stellar wind used in the
  simulations. 
\label{35v20n0.1}}
\end{figure}

\begin{figure}
\plotone{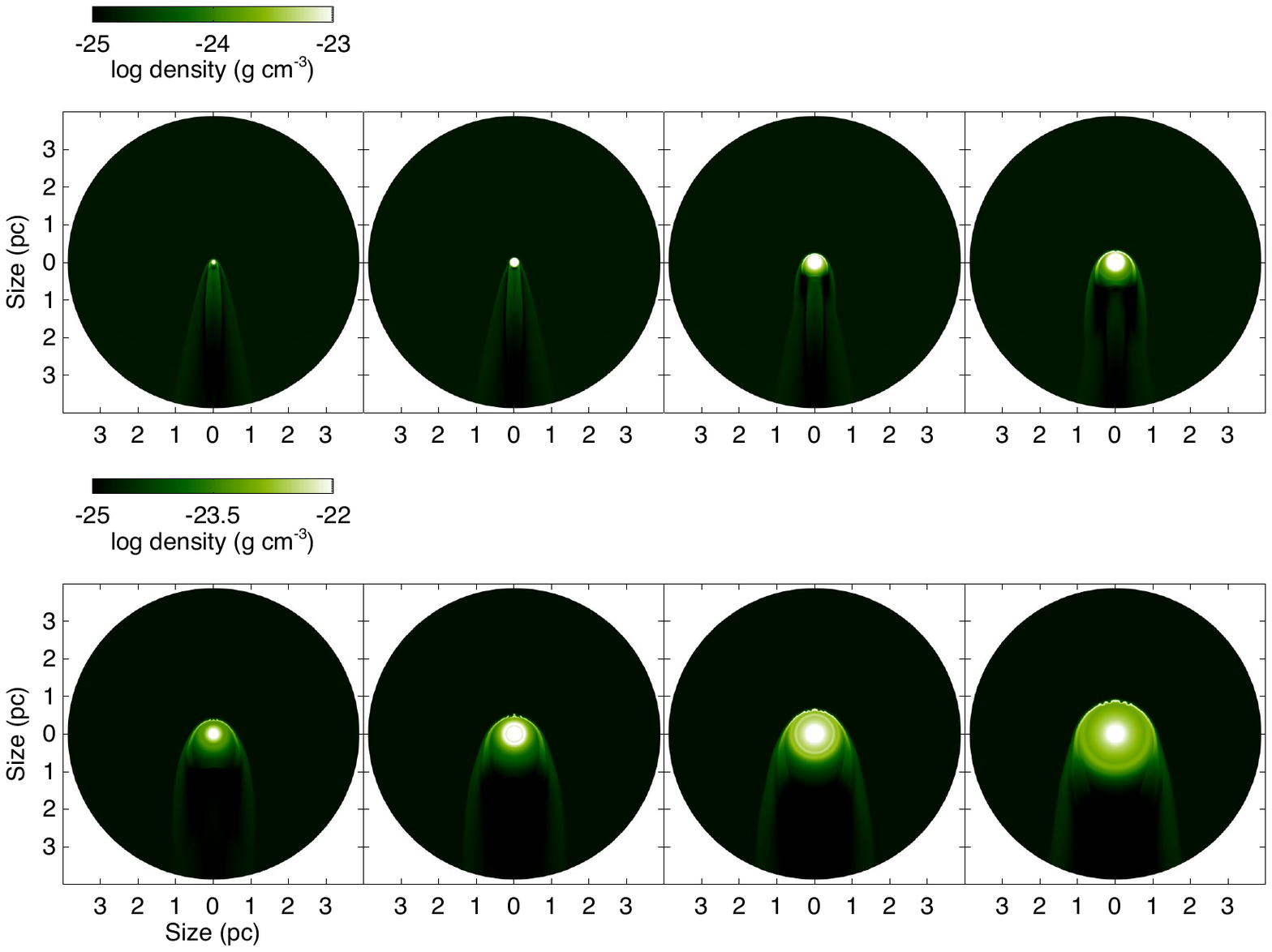}
\caption[ ]{Same as Fig.~\ref{35v20n0.1} but for a star moving at 50
  $\rm km\,s^{-1}$ through an ISM with density 0.1 ${\rm cm}^{-3}$.
\label{35v50n0.1}}
\end{figure}

In  Figs.~\ref{35v20n0.1} and \ref{35v50n0.1}  we show the evolution
of the gas density in log scale of an AGB star
with initial mass 3.5 \Mso.  We have run three simulations for this
stellar model (see 3 bottom lines of Table 1) testing two velocities
for the interaction: 20 (with an ISM density of 0.1\cm3),  and 50 \kms
(using ISM densities of 0.1 and 0.01\cm3) covering two orders of
magnitude in the value of the ram pressure. To study the process of
interaction of this more massive star we have used some of the
parameters already studied for the low mass star. In this way we better
isolate the effect of the stellar mass, as reflected in the mass-loss history,
in the simulations.

The first panel on the top left of the
figure has been selected at 
1.9 $\times 10^5~\mathrm{yr}$ into the evolution (see first mark in
Fig.~\ref{agbwindevo35}), and the subsequent panels from left to right and top
to bottom have been selected every 4.8$\times10^3~\mathrm{yr}$. We have
chosen not to show the last thousand years of the evolution given that
the shell morphology, besides growing in size, does not change
meaningfully from the last output (lower right corner) 
shown in Figs.~\ref{35v20n0.1} and \ref{35v50n0.1}. 

The CSE formed at the early AGB phase, during the
constant wind period, are identical to the ones developed for
the lower mass model, given that the inputs of the simulations in terms of
mass-loss rate are the same during the early AGB phase. The shape of the
CSE at this early stage depends then on the parameters assumed for
the interaction.

As expected, a bow-shock shape appears in the 
direction of the movement, but also a cometary tail is formed which is fed directly from
the stellar wind and from material stripped away from the bow
shock. The outermost shell departs from sphericity given that the
pressure encountered by the stellar wind is not isotropic, since the
stellar wind running into a higher pressure environment in the
opposite direction to the stellar movement. 

Once the mass-loss increases, it is kept continuously at high rates
until the tip of TP-AGB. The wind expands within the cavity carved by the
previously stellar wind during the whole AGB evolution of the
star. Material in the bow shock is regularly deflected downstream
of the movement and replenished by the stellar wind. The large amount of
material contained in the outermost shell can
compete effectively with the ISM ram pressure. Larger values of ram
pressure (ISM relative velocity) need to be used in order to
strip completely the outermost shell, and therefore, the freshly
ejected material from the star never encounters the ISM  directly. No 
pressure balance is reached between the CSE and the ISM. As a consequence the
shell continuously grows in the upstream direction. The wind expands in
the downstream direction thorough a lower pressure cavity, no shock
region develops and therefore no shell with bright edges is present.

The CSE  is able to grow in size faster in the upstream
direction for the same parameters of the interaction for the more
massive star. The mass-loss rate history adopted in our models 
is constrained by the stellar evolution models of \cite{Vw93}. There
is more than an order of magnitude difference in the maximum mass-loss
reached by the low and high mass stars considered ($0.5 \times 10^{-5}$ versus $0.2 \times
10^{-4} $\mlr for the 1 \Mso and 3.5 \Mso star respectively) when they
reach the tip of the AGB. In addition,
the mass-loss history is different, and so are the timescales of
the evolution of the star in the AGB.  The overall features of the 
interaction are the same as for the low-mass star, however, the details of the
interaction differ, stressing the importance of the
stellar input. 

The asymmetry of the
shell is barely detectable at the end of the AGB 
evolution in the R$_\mathrm{3.5M}$~20h model.  We define $\omega$ as
the ratio between the shell´s radius in the 
direction of the movement versus the radius in the perpendicular
direction, in order to quantify the effect of the interaction in the
morphology of the shell when comparing the two models with different
stellar masses.  We get a value of $\omega \approx 0.85$ for the
last $10^4$ yr of the AGB evolution. For comparison $\omega = 0.63$ for 
the 1 \Mso model with the same parameters for the interaction \citep{Vgm03}.

Despite the fact that relatively massive stars within the range of
AGB progenitors are not expected to have a
large velocity within the Galaxy, we have run simulations with
an intermediate relative velocity of 50 \kms. 
Model R$_\mathrm{3.5M}$~50h is shown in Fig.~\ref{35v50n0.1}.  
Instabilities appear in this model but do not lead to a complete
destruction of the shell as they do for the same parameters of the
interaction but a low mass star. 
At the end of the AGB evolution $\omega$ is 0.8 and so the disruption of
the morphology is significant. Instabilities appear 
but they do not have the radical influence
in the morphology that they have for the low mass model for the same parameters. 

A third model, for the 3.5 \Mso star, is listed in Table 1
R$_\mathrm{3.5M}$~50l. This model represents a low
density ISM (0.01 \cm3) with an intermediate relative velocity
of 50 \kms. The CSE has a similar evolution as the one described above
and at the end of the AGB we measure an $\omega =
0.95$. The deformity in the morphology of the shell at the end of the
AGB will be negligible. This is consistent with the strength of the parameters
assumed for the interaction given that this model has  a slightly lower ram pressure than the
model R$_\mathrm{3.5M}$~20h at a lower velocity. 

We find that for values of the product n$_{\rm ISM} v \approx$ 25-40
or ram pressures  4-6 $\times10^{-13}$ ${\rm dyn~cm^{-2}}$ 
the interaction shows clearly during most 
of the AGB phase, but disappears, or will be barely
noticeable during the last $10^4$ yr of the AGB evolution and surely
will not leave any imprint on the Planetary Nebulae morphology. 

\section{DISCUSSION}

The number of detected AGB stars with asymmetric shells is growing and
with it the evidence that the interaction between the circumstellar
envelope and the ISM is a common phenomena.  Our models show that this
is indeed the case and that the morphological features detected so far
agree very well with the models presented here. Examples of the morphology of AGB stars 
interacting with the ISM can be found in different wavelengths and thus
trace both the gas and the dust components: in CO and HI (see
e.g. \citealt{Mr07,Mat08,Lib08,Lib09,Lib10,Mat11}), in UV
\citep{Metal07,Sah10}, in the optical \citep{Fu10}, and in the IR 
\citep{Lad10,Metal11,Uetal06,U08,Uetal10, Jor11}. 

A new wealth of information soon will be
available on the distribution of dust around AGB stars \citep{Gro11}
and with it the possible detection of more asymmetric shells caused
by the interaction with the ISM. \cite{vanetal11} shows in 
a simulation of a fast moving red supergiant star that the dust
distribution follows  that of the gas when the grains considered are
small, and that bigger dust grains penetrate further into the unshocked ISM.
In general, for AGB stars, the size distribution of dust grains is, in
principle, smooth (see e.g. \citealt{nibita07}) with the stars
in the lower mass range having small amounts of  dust. No size
segregation is, in principle, expected  from different progenitor
masses or dust chemical composition (C versus O-rich dust). Dust
processing, however, is envisaged as the ejected AGB material mixes
with the ISM non-processed matter \citep{Vgm02,Vmg02}, as well as changes from
amorphous to crystalline as the star evolves from the AGB to the PN
phase \citep{Stan07}. Small grains are more
abundant and have a larger collective radiative surface area thus
dominating the infrared emission \citep{vanetal11}. Therefore
gas distributions presented in this paper should trace the
dust distribution from infrared observations closely.
Moreover, in the few cases
where the emission at different wavelengths can be compared
\citep{Rl09,Metal11} it seems that the distribution of dust and gas are
fairly well coupled.

According to our simulations, stars observed during the early AGB phase, where a
constant low density wind can be assumed, are all expected to show bow
shock structures characteristic of the interaction with the ISM. This holds true 
for the range of conditions
explored for the interaction  (expanding two orders of magnitude in the value assumed for
the ram pressure), and for the range of the stellar
masses considered (1 and 3.5 \Mso). If this early AGB wind lasts 
long enough to reach pressure equilibrium with the ISM, the location of the bow shock
can be estimated analytically.

Our low velocity models for the interaction show that, as the
star ascends toward the tip of the AGB on the HR diagram, the stellar wind always expands
within the cavity created by the previous stellar wind.  However, even
in this case, the AGB structure is very much influenced. As already
pointed out in \cite{Vgm03}, ram pressure stripping operates very
efficiently in the shell interacting directly with the ISM, reducing
substantially the mass of the envelope. Most the stellar mass lost
along the AGB is not to be found in the CSE, ejected mass is
continuously removed and left behind the star.
This is in contrast with the expectations for an AGB star at rest. 
If the star does not move relatively to its local ISM, the mass
in the CSE will contain the amount of mass lost by the star plus
a non-negligible amount of ISM mass swept out by the wind 
(see i.e. \citealt{Vgm02}). 

We gather that the influence of the
interaction in the morphology strongly depends on the time at which
the star is observed along the evolution. Early-AGB stars all show
very asymmetric bow shocks along the direction of the stellar motion
and collimated tails in the opposite direction. The asymmetry due to the
interaction is maintained, and will be detectable, along most of
the AGB for all the ram pressure conditions and stellar masses used 
in this work. At the end of the AGB, however, after most of the stellar
mass is lost, a shell grows in size that is capable of competing with the ram
pressure provided by the ISM when the stellar velocity is low (10
\kms) and the stellar mass is large (and with it the mass-loss).
In the Galaxy, the different stellar components are characterized by 
different dynamical behavior. Disk stars have velocities of 10-40 
\kms and are not expected to be found far above the galactic plane. If 
we assume that the ISM gas and the stars are not moving exactly at 
the same speeds within the Galaxy, it is safe to expect that disk stars could be moving
with respect to the ISM at low velocities (10-20 \kms) or, for some
extreme cases at intermediate velocities (30-50 \kms). 

To test the validity of the application of our models we can compare
the obtained physical sizes with observed AGB stars 
with well determine distances: CW Leonis (Ladja et al. 2010), 
IRC+10216 (Sahai \& Chronopoulus 2010) and Mira (Martin et al. 2007).
CW Leonis has a typical bow shock structure at a distance of 0.26 pc
from the AGB star. This compares quite well with Fig. 11, at a time of
3.1 $\times$ 10$^5$ yr along the AGB, (ISM density of 0.1 cm$^{-3}$ ,
and v= 20 \kms). IRC+10216 has a typical bow shock structure at a
distance of 0.36 pc from the star. This compares well with Fig.
4 at 3.4 $\times$ 10$^5$ yr along the AGB, (ISM density of 1 cm$^{-3}$ ,
and v= 10 \kms). Mira shows an elongated tail of 4 pc. This compares very well with
Fig. 10; 1.5 $\times$ 10$^5$ yr along the AGB, (ISM density of 0.01
cm$^{-3}$ , and v= 100 \kms).

Regarding different stellar progenitors, our simulations show that the
effects  of the interaction are not noticeable in the
morphology of the shells at the end of the AGB evolution of the 3.5
\Mso star unless we consider very high values of the ISM ram pressure.
The interaction, however, reduces the mass and the size of the
expected shell significantly.

\subsection{Stellar mass-loss on the AGB}
Although a full discussion regarding the treatment of mass-loss
during the AGB in stellar evolution calculations is beyond the scope of this work, 
it is in place to deliberate about the choice of the stellar evolution calculations 
used in our models. 

Mass-loss is a crucial process in the evolution of stars along the AGB,
however, it cannot be calculated from first principles. High mass-loss 
rates (the so-called superwind (SW); \citealt{Ren81}) are needed to remove the stellar envelope
at the tip of the AGB. The scenario generally accepted nowadays for the development of these 
high mass-loss rates involves two entangled processes: 
shock waves caused by the stellar pulsation and the acceleration of dust by radiation 
pressure. The stellar pulsation in AGB stars creates shock waves
which propagate through the stellar atmosphere. The dissipation of the
mechanical energy associated with these shocks leads to the levitation of the
upper layers of the atmosphere, where the gas becomes sufficiently cool (by
expansion and by dilution of the stellar radiation field) and dense
to allow heavy elements to condense into grains.
As grains nucleate and grow they
experience the force exerted by the stellar radiation pressure and thus
are accelerated. The momentum coupling between gas and dust drives
the outflow. Pioneering work on dynamical calculations to drive the stellar
wind were made by \cite{Wood79} and \cite{Bow}.

The calculation of
mass-loss during the AGB requires using dynamical model atmospheres in which 
time-dependent dynamics (shock waves and
winds), radiation transfer (strong variable stellar radiation field), and dust
and molecular formation processes need to be
considered together. Several calculations of dust-driven winds on the AGB are 
available in the literature. \cite{Wds94} models give mass-loss
rates as a function of the fundamental stellar parameters for stationary
atmospheres. Dynamical model atmospheres calculations have been computed 
by \cite{Bow,Fgs92,Arn,Hof98}. Given computational limitations and our current
knowledge of the fundamental physical data, becomes clear why a full reliable 
prescription for mass-loss on the AGB is still not available in the literature 
(see e.g. \citealt{Will00,Will08}). 
In particular, \cite{Hof98} demonstrated how changes in the
micro-physics result in considerable different mass-loss rates. 

Stellar evolutionary models
follow the temporal behavior of the mass-loss during the AGB
\citep{Vw93,Blo95,Sch99,Sc05,Wach02} under different prescriptions. 
Although the mass-loss rates are not derived from first principles in these
models, and most of them rely either on the dynamical model atmosphere 
calculations of \cite{Bow} and
\cite{Arn}, or on the semi-empirical mass-loss rates formula
derivations of \cite{Wood90}, they do provide a unique opportunity to study the
extensive history of mass-loss on the AGB and beyond.
The comparison of the different mass-loss prescriptions with observations 
of individual stars in the AGB evolutionary phase is complicated 
given the variable nature of the star (and the mass-loss), and a full discussion on the subject would be 
lengthy and beyond the scope the scope of this work. However, it is important to note that the
\cite{Vw93} prescription is the most widely used and not only has not been 
ruled out by observations, but even favored over other parameterizations (see e.g. 
\citealt{Zbm02}).

In our simulations, after the early AGB phase, the star ascends the termal-pulsing AGB where
the wind is characterized by a series of superwind events modulated by
the thermal pulses. We have modeled the
evolution of the star along this phase by constraining the stellar
wind input in the simulations the \cite{Vw93} parameterization. It is important to note that the
maximum mass-loss rate adopted in the \cite{Vw93}
models is limited to the radiation pressure limit. Dust driven wind models
allow values to be up to two times this limit. If the maximum
mass-loss rates increase, the timescale of the evolution of the star is modified
accordingly. By considering the evolution 
of the mass-loss evolution provided by a set stellar evolution calculations we 
are simulating a realistic scenario with no free parameters (note that the mass-loss 
in other studies is arbitrarily chosen to possible AGB values). 

We have taken a
particular set of stellar evolution 
models and simulate the effects of the stellar motion in the
shell formation along the AGB phase. The fine details of the evolution
are expected to change under other prescriptions for the evolution of
the star, however, the bulk of the results will remain
for the most part unchanged. A variable wind scenario is expected for pulsating AGB stars 
in the thermal pulsing phase. There are two main differences between the models presented 
here and the compilation by \cite{War07}: i) the use of variable winds and different 
stellar masses to describe the AGB evolution of the star, and ii) the use of a cooling function
under $10^4$ \,K and realistic wind temperatures. Although these models represent a fair 
simplification of the problem, a more realistic description of the interaction requires the use 
of variable winds and realistic cooling functions.

\section{CONCLUSIONS}
All models, independent of the parameters used for the interaction, show prominent 
bow-shock structures during the early-AGB evolution. This phase is characterized by 
constant mass-loss rates of the order of $\sim 10^{-8}$ \Mso that reach pressure 
equilibrium with the ISM. The morphology of the shell is a direct translation of the 
strength of the interaction. 

Models representing a population of stars moving with relative speeds at 
the lower end of the velocity dispersion are characteristic of disk stars (10 to 20 \kms) all 
and show the same pattern. The wind always expands within the
bow-shock cavity and because it is  
not interacting directly with the ISM the instabilities do not grow. Furthermore, at the
tip of the AGB the stellar wind can compete with the ram pressure provided by the ISM and 
the shell grows in size in the upstream direction. The fact that the wind always encounters 
a smaller pressure in the opposite direction of the motion causes the star to be 
displaced with respect the geometrical center of the envelope.

Regarding the higher end of the velocity dispersion of the disk population (30 to 50 \kms) 
we show that, as ram pressure stripping becomes more efficient, the stellar wind interacts 
directly with the ISM as the thermal-pulses take place. The main consequence is the formation 
of instabilities in the shells. While in the upstream direction,
the AGB wind interacts directly with the ISM leading to
the formation of a bow-shock, in the downstream direction, the wind expands
within the tunnel left behind by the star. Prominent elongated tails are formed in the
downstream direction for these models. 

We find that under the velocities expected for a population of halo stars the 
circumstellar envelopes are expected to be instable and get fragmented. 
Mixing of the AGB wind with ISM material is dominant as the ISM penetrates further 
as the bow-shock breaks up. The appearance of oblique shocks when the radiative cooling 
is stronger (i.e. higher ISM densities) exacerbates the fragmentation effect.

The observable effects of the interaction, although clearly visible,  along most 
of the evolution along the AGB they disappear in the morphological features at the
end of the evolution for the more massive, 3.5 \Mso models. The
interaction is not expected to leave any imprint on the Planetary
Nebulae morphology for these models. 

In general, ram pressure stripping is a powerful mechanism to remove mass from the 
envelopes of evolved stars, and even though in some cases the interaction is not 
expected to leave an imprint on the morphology of the envelope, its effects are 
important in reducing the mass and the size with respect to the values expected 
if the star had zero relative velocity with respect to the ISM.

\acknowledgments 
We thank the anonymous referee for very relevant comments that resulted 
in an improvement of the original version of this paper.

\begin{deluxetable}{lrrrccrc}
\tabletypesize{}
\tablecaption{Model parameters}
\tablewidth{0pt}
\tablehead{
\colhead{Run ID} &
\colhead{v$_{\rm ISM}$} &
\colhead{n$_{\rm ISM}$} & 
\colhead{T$_{\rm ISM}$} & 
\colhead{P$_{\rm RAM}$} & 
\colhead{Mach$_{\rm ISM}$} &
\colhead{$r_\mathrm{so}$}  &
\colhead{} \\
\colhead{ }   & 
\colhead{[\,km\,s$^{-1}$]} &
\colhead{[${\rm cm}^{-3}$]} &
\colhead{[${\rm K}$}] &
\colhead{[${\rm dyn~cm^{-2}}$]}&
\colhead{} &
\colhead{[$\times 10^{17}{\rm cm}$]} &
\colhead{} 
}
\startdata

R$_\mathrm{1M}$~10h&     10 & 1     & 100& 1.67$\times$10$^{-12}$ & 8.5 &0.77&\\
R$_\mathrm{1M}$~10l&     10 & 0.1  & 100 & 1.67$\times$10$^{-13}$ & 8.5&2.46&\\
R$_\mathrm{1M}$~20h&     20 & 0.1  &6000&  6.69$\times$10$^{-13}$& 2.2 &1.23&\\
R$_\mathrm{1M}$~20l&     20 & 0.01  &6000&  6.69$\times$10$^{-14}$ & 2.2&3.89&\\
R$_\mathrm{1M}$~30l&     30 & 0.1   & 6000& 1.33$\times$10$^{-12}$ & 3.4&0.79&\\
R$_\mathrm{1M}$~50h&     50 & 0.1  &6000&4.18$\times$10$^{-12}$ & 5.5 &0.49&\\
R$_\mathrm{1M}$~50l&     50 & 0.01  &6000&  4.18$\times$10$^{-13}$ & 5.5&1.56&\\
R$_\mathrm{1M}$~85l&    85  & 0.05  &6000&6.04$\times$10$^{-12}$ & 9.3 &0.41&\\
R$_\mathrm{1M}$~100h&   100 & 0.1  &6000&1.67$\times$10$^{-11}$ &11.0&0.25&\\
R$_\mathrm{1M}$~100l&   100 & 0.01  &6000&1.67$\times$10$^{-12}$ &11.0& 0.79&\\
\hline
R$_\mathrm{3.5M}$~20h&   20 & 0.1  &6000&  6.69$\times$10$^{-13}$ & 2.2&1.34&\\
R$_\mathrm{3.5M}$~50h&   50 & 0.1  &6000&  4.18$\times$10$^{-12}$ & 5.5&0.54&\\
R$_\mathrm{3.5M}$~50l&   50 & 0.01  &6000&  4.18$\times$10$^{-13}$ & 5.5&1.72&\\
\hline
\enddata
\label{tableruns}
\end{deluxetable}

\begin{deluxetable}{lclcrcr}
\tabletypesize{}
\tablecaption{Comparisson between models under the parameters for Mira}
\tablewidth{0pt}
\tablehead{
\colhead{Reference} &
\colhead{v$_{\rm ISM}$} &
\colhead{n$_{\rm ISM}$} & 
\colhead{T$_{\rm ISM}$} & 
\colhead{$\dot{M}_{\rm wind}$} & 
\colhead{v$_{\rm wind}$} &
\colhead{T$_{\rm wind}$} \\
\colhead{ }   & 
\colhead{[\,km\,s$^{-1}$]} &
\colhead{[${\rm cm}^{-3}$]} &
\colhead{[${\rm K}$}] &
\colhead{[$10^{-7}$ M$_\odot$~yr$^{-1}$]}&
\colhead{[\,km\,s$^{-1}$]} &
\colhead{[${\rm K}$]} }
\startdata
Model R$_\mathrm{1M}$~100l & 100 & 0.01 & 6\,000 & 0.1--1.4& 2--7& 2\,700\\
Model R$_\mathrm{1M}$~100h &     & 0.1  &   &    &      &                          \\
Wareing et al.(2007b)&  130 & 0.03 & 8\,000 & 1-3    & 5 & 10\,000 \\
Esquivel et al.(2010)& 125 & 1    & 1\,000 & 3           & 5 & 100  \\
                    &     & 0.05 & $10^6$ &              &   &  \\

\hline
\enddata
\label{tableruns}
\end{deluxetable}

\end{document}